\documentclass[11pt]{article}

\usepackage[hyphens]{url}  
\usepackage{graphicx} 
\usepackage{natbib}  
\usepackage{caption} 
\usepackage[utf8]{inputenc}
\usepackage{amsmath}
\usepackage{amsthm}
\usepackage{amsfonts}
\usepackage{amssymb}
 \usepackage{fullpage}
\usepackage{mathtools}
\usepackage{thm-restate}
\usepackage{mleftright}
\usepackage{authblk}
\usepackage{bm}
\usepackage[linesnumbered,ruled,lined,noend]{algorithm2e}
\usepackage{hyperref}
\usepackage[capitalize,noabbrev]{cleveref}
\usepackage[table]{xcolor}
\usepackage{etoolbox}
\usepackage{xparse}
\usepackage{nicefrac}
\usepackage{booktabs}
\usepackage{caption}
\usepackage{enumitem}
\usepackage{float}
\usepackage{array}
\usepackage{pifont}
\usepackage{mdframed}
\usepackage{blkarray}

\usepackage{bbm}

\newcommand{\declarecolor}[2]{\definecolor{#1}{RGB}{#2}\expandafter\newcommand\csname #1\endcsname[1]{\textcolor{#1}{##1}}}
\declarecolor{White}{255, 255, 255}
\declarecolor{Black}{0, 0, 0}
\declarecolor{Maroon}{128, 0, 0}
\declarecolor{Coral}{255, 127, 80}
\declarecolor{Red}{182, 21, 21}
\declarecolor{LimeGreen}{50, 205, 50}
\declarecolor{DarkGreen}{0, 100, 0}
\declarecolor{Navy}{0, 0, 128}
\hypersetup{
	colorlinks=true,
	pdfpagemode=UseNone,
	citecolor=DarkGreen,
	linkcolor=Navy,
	urlcolor=Navy,
	pdfstartview=FitW
}

\newcommand{\seq}[1]{\texttt{#1}}

\usepackage{tikz}
\usetikzlibrary{arrows.meta,calc,shapes.misc,arrows}
\tikzset{cross/.style={path picture={
  \draw[black]
(path picture bounding box.south east) -- (path picture bounding box.north west) (path picture bounding box.south west) -- (path picture bounding box.north east);
}}}
\tikzstyle{chanode}=[fill=white,draw=black,circle,cross,inner sep=1mm]
\tikzstyle{pl1node}=[fill=black,draw=black,circle,inner sep=.8mm]
\tikzstyle{pl2node}=[fill=white,draw=black,circle,inner sep=.8mm]
\tikzstyle{termina}=[fill=white,draw=black,inner sep=.8mm]

\usetikzlibrary{fit,shapes.misc,arrows.meta}
\makeatletter
\tikzset{
  fitting node/.style={
    inner sep=0pt,
    fill=none,
    draw=none,
    reset transform,
    fit={(\pgf@pathminx,\pgf@pathminy) (\pgf@pathmaxx,\pgf@pathmaxy)}
  },
  reset transform/.code={\pgftransformreset}
}
\tikzset{cross/.style={path picture={
  \draw[black]
(path picture bounding box.south east) -- (path picture bounding box.north west) (path picture bounding box.south west) -- (path picture bounding box.north east);
}}}
\tikzstyle{ox}=[semithick,draw=black,circle,cross,inner sep=1.2mm]
\makeatother

\newcommand{\depth}{\mathfrak{D}}

\crefname{assumption}{assumption}{assumptions}

\newcommand{\etatri}{\eta_{\triangle}}
\newcommand{\uttri}{\vec{u}_{\triangle}}

\newcounter{qst}
\crefname{qst}{Question}{Questions}

\input{figs/tikzmacros.tex}

\theoremstyle{plain}
\newtheorem{theorem}{Theorem}[section]
\newtheorem{lemma}[theorem]{Lemma}
\newtheorem{corollary}[theorem]{Corollary}
\newtheorem{proposition}[theorem]{Proposition}

\newtheorem{claim}[theorem]{Claim}

\theoremstyle{definition}
\newtheorem{definition}[theorem]{Definition}

\theoremstyle{remark}
\newtheorem{remark}[theorem]{Remark}

\newcommand{\fp}{\textsc{FixedPoint}}

\newcommand{\obsutil}{\textsc{ObserveUtility}}
\newcommand{\nextstrat}{\textsc{NextStrategy}}

\DeclareMathOperator{\reg}{Reg}

\title{Near-Optimal $\Phi$-Regret Learning in Extensive-Form Games}

\author[1]{Ioannis Anagnostides}
\author[2]{Gabriele Farina}
\author[3]{Tuomas Sandholm}

\affil[1,3]{Carnegie Mellon University}
\affil[2]{Meta AI}
\affil[3]{Strategy Robot, Inc.}
\affil[3]{Optimized Markets, Inc.}
\affil[3]{Strategic Machine, Inc.}
\affil[ ]{\texttt {\{ianagnos,gfarina\}@cs.cmu.edu}, and \texttt{sandholm@cs.cmu.edu}}

\date{}

\makeatletter
    \patchcmd\algocf@Vline{\vrule}{\vrule \kern-0.4pt}{}{}
    \patchcmd\algocf@Vsline{\vrule}{\vrule \kern-0.4pt}{}{}
\makeatother
\SetKwComment{Hline}{}{\vspace{-3mm}\textcolor{gray}{\hrule}\vspace{1mm}}
\definecolor{darkgrey}{gray}{0.3}
\definecolor{commentcolor}{gray}{0.5}
\SetKwComment{Comment}{\color{commentcolor}[$\triangleright$\ }{}
\SetCommentSty{}
\SetNlSty{}{\color{darkgrey}}{}
\SetKwProg{Fn}{function}{}{}
\SetKwProg{Subr}{subroutine}{}{}
\crefalias{AlgoLine}{line}%
\crefname{algocf}{Algorithm}{Algorithms}

\newcommand*{\N}{{\mathbb{N}}}
\newcommand*{\R}{{\mathbb{R}}}

\newcommand{\regdep}{\mathfrak{R}}

\newcommand{\mul}{\mu}
\newcommand{\ind}{r}
\newcommand{\emptyseq}{\varnothing}

\newcommand{\cR}{\mathcal{R}}
\newcommand{\cA}{\mathcal{A}}
\newcommand{\cQ}{\mathcal{Q}}
\newcommand{\cJ}{\mathcal{J}}
\newcommand{\cZ}{\mathcal{Z}}
\newcommand{\cH}{\mathcal{H}}

\newcommand{\tree}{\mathcal{T}}
\newcommand{\treeset}{\mathbb{T}}
\newcommand{\edg}{e}

\DeclareMathOperator{\polylog}{polylog}

\newcommand{\cX}{\mathcal{X}}
\newcommand{\defeq}{\coloneqq}

\newcommand{\algoshort}{\texttt{LRL-OFTRL}\xspace}
\newcommand{\oftrl}{\texttt{OFTRL}}
\DeclareMathOperator*{\argmax}{arg\,max}

\renewcommand{\vec}[1]{\bm{#1}}

\newcommand{\nakedcite}[1]{\citeauthor{#1}, \citeyear{#1}}

\newcommand{\range}[1]{[\![#1]\!]}
\newcommand{\ut}{\vec{u}}
\newcommand{\Ut}{\vec{U}}
\newcommand{\vx}{\vec{x}}
\newcommand{\vy}{\vec{y}}
\newcommand{\vxstar}{\vec{x}^\star}
\newcommand{\phistar}{\phi^\star}

\newcommand{\vq}{\vec{q}}
\newcommand{\vr}{\vec{r}}
\newcommand{\vb}{\vec{b}}
\newcommand{\vX}{\vec{X}}
\newcommand{\vlam}{\vec{\lambda}}

\newcommand{\vS}{\vec{S}}
\newcommand{\mat}[1]{\mathbf{#1}}

\def\[#1\]{%
  \begin{align*}%
    #1%
  \end{align*}%
}
\NewDocumentCommand{\numberthis}{om}{%
\IfNoValueTF{#1}{%
    \refstepcounter{equation}\tag{\theequation}%
  }{%
    \tag{#1}%
  }%
  \label{#2}%
}

\begin{document}

\maketitle
\pagenumbering{gobble}
\begin{abstract}
    In this paper, we establish efficient and uncoupled learning dynamics so that, when employed by all players in multiplayer perfect-recall imperfect-information extensive-form games, the \emph{trigger regret} of each player grows as $O(\log T)$ after $T$ repetitions of play. This improves exponentially over the prior best known trigger-regret bound of $O(T^{1/4})$, and settles a recent open question by~Bai et al. (2022). As an immediate consequence, we guarantee convergence to the set of \emph{extensive-form correlated equilibria} and \emph{coarse correlated equilibria} at a near-optimal rate of $\frac{\log T}{T}$. 

Building on prior work, at the heart of our construction lies a more general result regarding fixed points deriving from rational functions with \emph{polynomial degree}, a property that we establish for the fixed points of \emph{(coarse) trigger deviation functions}. Moreover, our construction leverages a refined \textit{regret circuit} for the convex hull, which---unlike prior guarantees---preserves the \emph{RVU property} introduced by~Syrgkanis et al. (NIPS, 2015); this observation has an independent interest in establishing near-optimal regret under learning dynamics based on a CFR-type decomposition of the regret.
\end{abstract}

\clearpage
\pagenumbering{arabic}

\section{Introduction}
\label{section:introduction}

A primary objective of artificial intelligence is the design of agents that can adapt effectively in complex and nonstationary multiagent environments---modeled as \emph{general-sum games}. Multiagent decision making often occurs in a decentralized fashion, with each agent only obtaining information about its own reward function, and the goal is to \emph{learn} how to play the game through repeated interactions. But \emph{how do we measure the performance of a learning agent?} A popular metric commonly used is that of \emph{external regret} (or simply regret). However, external regret can be a rather weak benchmark: a no-external-regret agent could still incur substantial regret under simple in-hindsight ``transformations'' of its behavior---\emph{e.g.}, consistently switching from an action $a$ to a different action $a'$~\citep{Gordon08:No}.

A more general metric is \emph{$\Phi$ regret}~\citep{Hazan07:Computational,Rakhlin11:Online,Stoltz07:Learning,Greenwald03:A}, parameterized by a set deviations $\Phi$. From a game-theoretic standpoint, the importance of this framework is that different choices of $\Phi$ lead to different types of equilibria~\citep{Greenwald03:A,Stoltz07:Learning}. For example, one such celebrated result guarantees that \emph{no-internal-regret} players converge---in terms of empirical frequency of play---to the set of \emph{correlated equilibria (CE)}~\citep{Foster97:Calibrated,Hart00:Simple}. This brings us to the following central question:
\begin{quote}
    \centering
    \emph{What are the best performance guarantees when no-$\Phi$-regret learners are playing in multiplayer general-sum games?}
\end{quote}
Special cases of this question have recently received considerable attention~\citep{Daskalakis11:Near,Rakhlin13:Online,Rakhlin13:Optimization,Syrgkanis15:Fast,Foster16:Learning,Wei18:More,Chen20:Hedging,Hsieh21:Adaptive,Daskalakis21:Near,Daskalakis21:Fast,Anagnostides21:Near,Piliouras21:Optimal}. In particular, \citet{Daskalakis21:Near} were the first to establish $O(\polylog T)$ external regret bounds for normal-form games,\footnote{With a slight abuse of notation, we use the $O(\cdot)$ notation in our introduction to suppress parameters that depend (polynomially) on the natural parameters of the game.} and subsequent work extended those results to internal regret~\citep{Anagnostides21:Near}; those guarantees, applicable when \emph{all} players employ specific learning dynamics, improve exponentially over what is possible when a player is facing a sequence of adversarially produced utilities---the canonical consideration in online learning. However, much less is known about $\Phi$-regret learning beyond normal-form games.

One such important application revolves around learning dynamics for \emph{extensive-form correlated equilibria (EFCE)}~\citep{Stengel08:Extensive,Gordon08:No,Celli20:No-Regret,Morrill21:Efficient,Anagnostides22:Faster,Morrill21:Hindsight,Bai22:Efficient,Song22:Sample}. Indeed, a particular instantiation of $\Phi$ regret, referred to as \emph{trigger regret}, is known to drive the rate of convergence to EFCE~\citep{Farina21:Simple}. Incidentally, minimizing trigger regret lies at the frontier of $\Phi$-regret minimization problems that are known to be computationally tractable in extensive-form games. In this context, prior work established $O(T^{1/4})$ per-player trigger regret bounds~\citep{Anagnostides22:Faster}, thereby leaving open the possibility of obtaining near-optimal rates for EFCE; that question was also recently posed by~\citet{Bai22:Efficient}.

\subsection{Our Contributions}

Our main contribution is to establish the first uncoupled learning dynamics with near-optimal per-player trigger regret guarantees:

\begin{theorem}[Informal; precise version in \Cref{cor:trigger}]
    \label{theorem:efce}
    There exist uncoupled and computationally efficient learning dynamics so that the trigger regret of each player grows as $O(\log T)$ after $T$ repetitions of play.
\end{theorem}

This improves exponentially over the $O(T^{1/4})$ bounds obtained in prior work~\citep{Celli20:No-Regret,Farina21:Simple,Anagnostides22:Faster}, and settles an open question recently posed by~\citet{Bai22:Efficient}. As an immediate consequence, given that trigger regret drives the rate of convergence to EFCE (\Cref{prop:efce-trigger}), we obtain the first near-optimal rates to EFCE.

\begin{corollary}
    There exist uncoupled and computationally efficient learning dynamics converging to EFCE at a near-optimal rate of $\frac{\log T}{T}$.
\end{corollary}

\paragraph{Overview of our techniques} Our construction leverages the template of~\citet{Gordon08:No} for minimizing $\Phi$ regret (\Cref{algo:Gordon}). In particular, we follow the regret decomposition approach of~\citet{Farina21:Simple} to construct an external regret minimizer \emph{for the set of deviations} corresponding to \emph{trigger deviation functions}. A key difference is that we instantiate each regret minimizer using the recent algorithm of~\citet{Farina22:Near}, namely $\algoshort$, which is based on \emph{optimistic follow the regularizer leader (OFTRL)}~\citep{Syrgkanis15:Fast} under logarithmic regularization; $\algoshort$ guarantees suitable \emph{RVU bounds}~\citep{Syrgkanis15:Fast} for each ``local'' regret minimizer.

To combine those local regret minimizers into a global one for the set of trigger deviations that still enjoys a suitable RVU bound, we provide a refined guarantee for the ``regret circuit'' of the convex hull (\Cref{prop:rvu-hull}), which ensures that \emph{the RVU property is preserved} along the construction. Incidentally, this simple observation can be used to obtain the first near-optimal regret guarantees for algorithms based on a CFR-type decomposition of the regret~\citep{Zinkevich07:Regret}; as such, \Cref{prop:rvu-hull} has an independent and broader interest.

The next key step relates to the behavior of the fixed points of trigger deviation functions. (Fixed points are at heart of all known constructions for minimizing $\Phi$ regret~\citep{Hazan07:Computational}; see~\Cref{algo:Gordon}.) More precisely, to convert the RVU property from the space of deviations to the actual space of the player's strategies, we show that it suffices that the fixed points deriving from trigger deviation functions can be expressed as a rational function with a polynomial degree (\Cref{theorem:rational}). Importantly, we prove this property for the fixed points of trigger deviation functions (\Cref{prop:fp-efce}), thereby leading to \Cref{theorem:efce}; the last part of our analysis builds on a technique developed for obtaining $O(\log T)$ swap regret in normal-form games~\citep{Anagnostides22:Uncoupled}, although (imperfect-information) extensive-form games introduce considerable new challenges, not least due to the combinatorial structure of trigger deviation functions. We also obtain slightly improved guarantees for extensive-form \emph{coarse} correlated equilibria (EFCCE)~\citep{Farina20:Coarse}, a relaxation of EFCE that is attractive due to its reduced per-iteration complexity compared to 
EFCE.

Finally, we support our theory (\Cref{theorem:efce}) by implementing our algorithm and evaluating its performance through experiments on several benchmark extensive-form games in \Cref{section:experiments}. 

\subsection{Further Related Work}
\label{sec:related}

The notion of $\Phi$ regret has received extensive attention as a solution concept in the literature since it strengthens and unifies many common measures of performance in online learning (\emph{e.g.}, see ~\citep{Hazan07:Computational,Rakhlin11:Online,Stoltz07:Learning,Greenwald03:A,Marks08:No,Piliouras22:Evolutionary,Fujii23:Bayes,Bernasconi23:Constrained}). This framework has been particularly influential in game theory given that no-$\Phi$-regret learning outcomes are known to converge to different equilibrium concepts, depending on the richness of the set of deviations $\Phi$. For example, when $\Phi$ includes \emph{all constant transformations}---reducing to external regret---no-regret learning outcomes are known to converge to \emph{coarse correlated equilibria (CCE)}~\citep{Moulin78:Strategically}, a relaxation of CE~\citep{Aumann74:Subjectivity}. Unfortunately, CCE is understood to be a rather weak equilibrium concept, potentially prescribing irrational behavior~\citep{Dekel90:Rational,Viossat13:No,Giannou21:Survival}. This motivates enlarging the set of deviaitons $\Phi$, thereby leading to stronger---and arguably more plausible---equilibrium concepts. Indeed, the framework of $\Phi$ regret has been central in the quite recent development of the first uncoupled no-regret learning dynamics for EFCE~\citep{Celli20:No-Regret,Farina21:Simple} (see also~\citep{Morrill21:Efficient,Morrill21:Hindsight,Zhang22:A}).\footnote{While there are other methods for efficiently computing EFCE~\citep{Dudik09:SamplingBased,Huang08:Computing}, approaches based on uncoupled no-regret learning typically scale significantly better in large games.}

Our paper lies at the interface of the aforedescribed literature with a recent line of work that strives for improved regret guarantees when specific learning dynamics are in place; this allows bypassing the notorious $\Omega(\sqrt{T})$ lower bounds applicable under an adversarial sequence of utilities~\citep{Cesa-Bianchi06:Prediction}. The later line of work was pioneered by~\citet{Daskalakis11:Near}, and has been thereafter extended along several lines~\citep{Rakhlin13:Online,Rakhlin13:Optimization,Syrgkanis15:Fast,Chen20:Hedging,Daskalakis21:Near,Daskalakis21:Fast,Piliouras21:Optimal,Yang22:T}, incorporating partial or noisy information feedback models~\citep{Foster16:Learning,Wei18:More,Hsieh22:No,Bai22:Near}, and more recently, general Markov games~\citep{Erez22:Regret,Zhang22:Policy}. For additional pointers, we refer the interested reader to the survey of~\citet{Li22:Survey}.

A key reference point for our paper is the work of~\citet{Anagnostides22:Faster}, which established $O(T^{1/4})$ trigger regret bounds through \emph{optimistic hedge}. Specifically, building on the work of~\citet{Chen20:Hedging}, they showed \emph{multiplicative stability} of the fixed points associated with EFCE. While those works operate in the full information model, recent papers have also developed dynamics converging to EFCE under bandit feedback~\citep{Bai22:Efficient,Song22:Sample}. Finally, it is worth noting that $O(\polylog T)$ regret bounds in extensive-form games were already obtained in prior work~\citep{Farina22:Kernelized}, but they only applied to the weaker notion of external regret.
\section{Preliminaries}
\label{section:prel}

In this section, we introduce our notation and basic background on online learning and extensive-form games. For a more comprehensive treatment on those subjects, we refer to the excellent books of~\citet{Cesa-Bianchi06:Prediction} and~\citet{Leyton08:Essentials}, respectively.

\paragraph{Notation} We denote by $\N = \{1, 2, \dots \}$ the set of natural numbers. We use the variable $i$ with a subscript to index a player, and $t$ with a superscript to indicate the (discrete) time. To access the $\ind$-th coordinate of a $d$-dimensional vector $\vx \in \R^d$, for some index $\ind \in \range{d} \defeq \{1, 2, \dots, d\}$,
we use the symbol $\vx[\ind]$.

\subsection{Optimistic Online Learning and Regret}
\label{sec:regret}

Let $\cX \subseteq [0,1]^d$ be a nonempty convex and compact set, for $d \in \N$. In the framework of online learning, a learner (or a player), denoted by $\regdep$, interacts with the environment at time $t \in \N$ via the following subroutines.

\begin{itemize}
    \item $\regdep.\nextstrat()$: The learner outputs its next strategy $\vx^{(t)} \in \cX$ based on its internal state; and
    \item $\regdep.\obsutil(\ut^{(t)})$: The learner receives a feedback from the environment in the form of a utility vector $\ut^{(t)} \in\R^{d}$. 
\end{itemize}
The canonical measure of performance in online learning is the notion of \emph{regret}, denoted by $\reg^T$, defined for a fixed time horizon $T \in \N$ as
\begin{equation}
    \label{eq:extreg}
    \max_{\vxstar \in \cX} \left\{ \sum_{t=1}^T \langle \vxstar, \ut^{(t)} \rangle \right\} - \sum_{t=1}^T \langle \vx^{(t)}, \ut^{(t)} \rangle.
\end{equation}
In words, the performance of the learner is compared to the performance of playing an optimal \emph{fixed strategy} in hindsight. We will say that the agent has \emph{no-regret} if $\reg^T = o(T)$, under \emph{any} sequence of observed utilities.

\paragraph{Optimistic FTRL} By now, it is well-understood that broad families of online learning algorithms---such as \emph{follow the regularized leader}---incur at most $O(\sqrt{T})$ regret, even when the sequence of utilities is selected adversarially~\citep{Cesa-Bianchi06:Prediction}. In addition, significant improvements are possible when the observed utilities satisfy further properties, such as small variation ~\citep{Rakhlin13:Online}, which turns out to be crucial in the context of learning in games. To leverage such structure, \citet{Syrgkanis15:Fast} introduced \emph{optimistic follow the regularized leader (\oftrl)}, which updates every strategy $\vx^{(t+1)}$ as the (unique) solution to the optimization problem
\begin{equation}
    \label{eq:OFTRL}
    \tag{OFTRL}
    \max_{\vx \in \cX} \left\{ \left\langle \vx, \ut^{(t)} +  \sum_{\tau=1}^t \ut^{(\tau)} \right\rangle - \frac{1}{\eta} \cR(\vx) \right\},
\end{equation}
where $\eta > 0$ is the \emph{learning rate} and $\cR : \cX \to \R$ is a $1$-strongly convex regularizer with respect to some norm $\|\cdot\|$. \citet{Syrgkanis15:Fast} showed that $\oftrl$ satisfies a remarkable regret bound coined the \emph{RVU property}.

\begin{definition}[RVU property]
    \label{def:RVU}
    A regret minimizer satisfies the RVU property w.r.t. a dual pair of norms $(\| \cdot\|, \|\cdot\|_*)$ if there exist $\alpha, \beta, \gamma > 0$ such that for any $(\ut^{(t)})_{1 \leq t \leq T}$,
    \begin{equation*}
        \reg^T \leq \alpha + \beta \sum_{t=1}^{T-1} \|\ut^{(t+1)} - \ut^{(t)} \|_*^2 - \gamma \sum_{t=1}^{T-1} \|\vx^{(t+1)} - \vx^{(t)} \|^2.
    \end{equation*}
\end{definition}

\paragraph{$\Phi$ regret} A much more general performance metric than~\eqref{eq:extreg} is \emph{$\Phi$ regret}, parameterized by a set of \emph{transformations} $\Phi : \cX \to \cX$. Namely, $\Phi$-regret $\reg^T_{\Phi}$---for a time horizon $T \in \N$---is defined as
\begin{equation}
    \label{eq:phireg}
    \sup_{\phistar \in \Phi} \left\{ \sum_{t=1}^T \langle \phistar(\vx^{(t)}), \ut^{(t)} \rangle \right\} - \sum_{t=1}^T \langle \vx^{(t)}, \ut^{(t)} \rangle.
\end{equation}
External regret~\eqref{eq:extreg} is simply a special case of~\eqref{eq:phireg} when $\Phi$ includes all possible \emph{constant transformations}, but $\Phi$ regret can be much more expressive. A celebrated game-theoretic motivation for $\Phi$ regret stems from the fact that when all players employ suitable $\Phi$-regret minimizers, the dynamics converge to different notions of \emph{correlated equilibria}, well-beyond \emph{coarse correlated equilibria}~\citep{Foster97:Calibrated,Stoltz07:Learning,Hart00:Simple,Celli20:No-Regret}.

\paragraph{From external to $\Phi$ regret} As it turns out, there is a general template for minimizing $\Phi$ regret due to~\citet{Gordon08:No}. In particular, their algorithm assumes access to the following.

\begin{enumerate}
    \item \label{item:phiexternal} A \emph{no-external-regret} minimizer $\regdep_\Phi$ operating \emph{over the set of transformations} $\Phi$; and
    \item A \emph{fixed point oracle} $\fp(\phi)$ that, for any $\phi \in \Phi$, computes a fixed point $\vx \in \cX$, under the assumption that such a point indeed exists.\label{item:fp}
\end{enumerate}
Based on those ingredients, \citet{Gordon08:No} were able to construct a regret minimization algorithm $\regdep$ with sublinear $\Phi$ regret, as illustrated in \Cref{algo:Gordon}. Specifically, $\regdep$ determines its next strategy by first obtaining the strategy $\phi^{(t)}$ of $\regdep_{\Phi}$ (\Cref{line:stratphi}), and then outputting any fixed point of $\phi^{(t)}$ (\Cref{line:fp}). Then, upon receiving the utility vector $\ut^{(t)} \in \R^d$, $\regdep$ forwards as input to $\regdep_{\Phi}$ the utility function $\phi \mapsto \langle \ut^{(t)}, \phi(\vx^{(t)}) \rangle$. We will assume that $\Phi$ contains \emph{linear transformations}, in which case that utility can be represented as $\Ut^{(t)} \defeq \ut^{(t)} \otimes \vx^{(t)} \defeq \ut^{(t)} (\vx^{(t)})^\top \in \R^{d \times d}$ (\Cref{line:outer}); that is, $\otimes$ denotes the outer product of the two vectors. This algorithm enjoys the following guarantee.

\begin{algorithm}[!ht]
    \SetAlgoLined
    \SetInd{2.3mm}{2.3mm}
    \DontPrintSemicolon
    \KwData{An external regret minimizer $\regdep_{\Phi}$ for $\Phi$}    
        \Fn{$\nextstrat()$}{
            $\phi^{(t)} \leftarrow \regdep_{\Phi}.\nextstrat()$ \label{line:stratphi} \\
            $\vec{x}^{(t)} \leftarrow \fp(\phi^{(t)})$ \label{line:fp} \\
            \textbf{return} $\vec{x}^{(t)}$
        }
        \Hline{}
        \Fn{$\obsutil(\ut^{(t)})$}{
            Construct the utility $\Ut^{(t)} \leftarrow \ut^{(t)} \otimes\vec{x}^{(t)}$ \label{line:outer} \\
            $\regdep_{\Phi}.\obsutil(\Ut^{(t)})$
        }
    \caption{$\Phi$-Regret Minimizer~\citep{Gordon08:No}}
    \label{algo:Gordon}
\end{algorithm}

\begin{theorem}[\nakedcite{Gordon08:No}]
    \label{theorem:gordon}
    Let $\reg^T$ be the external regret of $\regdep_{\Phi}$, and $\reg_{\Phi}^T$ be the $\Phi$-regret of $\regdep$. Then, for any $T \in \N$,
    \begin{equation*}
        \reg^T = \reg_{\Phi}^T.
    \end{equation*}
\end{theorem}

It is also worth pointing out that a similar guarantee applies even under approximate fixed-point computations, as long as the accuracy is high enough~\citep{Gordon08:No}.

\paragraph{No-Regret learning in games} 
The main focus of our paper is about the behavior of no-regret learning dynamics when employed by all players in $n$-player games. More precisely, the strategy set of each player $i \in \range{n}$ is a nonempty convex and compact set $\cX_i$. Further, the utility function $u_i : \bigtimes_{i'=1}^n \cX_{i'} \to \R$ of each player $i \in \range{n}$ is multilinear, so that for any $\vx_{-i} \defeq (\vx_1, \dots, \vx_{i-1}, \vx_{i+1}, \vx_n)$, $u_i(\vx) \defeq \langle \vx_i, \ut_i (\vx_{-i}) \rangle$. 
%

In this context, learning procedures work as follows. At every iteration $t \in \N$ each player $i \in \range{n}$ commits to a strategy $\vx_i^{(t)} \in \cX_i$, and subsequently receives as feedback the utility corresponding to the other players' strategies at time $t$: $\ut_i^{(t)} \defeq \ut_i(\vx_{-i}^{(t)})$. It is assumed that players use no-regret learning algorithms to adapt to the behavior of the other players, leading to \emph{uncoupled} learning dynamics, in the sense that players do not use information about other players' utilities~\citep{Hart00:Simple,Daskalakis11:Near}. For convenience, and without any loss, we assume that $\| \ut_i^{(t)}\|_\infty \leq 1$, for $i \in \range{n}$ and $t \in \N$.

\subsection{Background on EFGs}

An \emph{extensive-form game (EFG)} is played on a rooted and directed tree with node-set $\cH$. Every \emph{decision (non-terminal) node} $h \in \cH$ is uniquely associated with a player who selects an action from a finite and nonempty set $\cA_h$. By convention, the set of players includes a fictitious ``chance'' player $c$ that acts according to a fixed distribution. The set of \emph{leaves (terminal) nodes} $\cZ \subseteq \cH$ corresponds to different outcomes of the game. Once the game reaches a terminal node $z \in \cZ$, every player $i \in \range{n}$ receives a payoff according to a (normalized) utility function $u_i : \cZ \to [-1, 1]$. 

In an imperfect-information EFG, the decision nodes of each player $i \in \range{n}$ are partitioned into \emph{information sets} $\cJ_i$, inducing a partially ordered set $(\cJ_i, \prec)$. For an information set $j \in \cJ_i$ and an action $a \in \cA_j$, we let $\sigma \defeq (j,a)$ be the \emph{sequence} of $i$'s actions encountered from the root of the tree until (and including) action $a$; we use the special symbol $\emptyseq$ to denote the \emph{empty sequence}. The set of $i$'s sequences is denoted by $\Sigma_i \defeq \{ (j,a) : j \in \cJ_i, a \in \cA_j \} \cup \{ \emptyseq \}$. We also let $\Sigma_i^* \defeq \Sigma_i \setminus \{\emptyseq\}$ and $\Sigma_j \defeq \{ \sigma \in \Sigma_i: \sigma \succeq j \}$, where we write $\sigma \succeq j$ if sequence $\sigma$ must pass from some node in $j$. We will use $\sigma_j$ to represent the \emph{parent sequence} of an information set $j \in \cJ_i$; namely, the last sequence before reaching $j$, or $\emptyseq$ if $j$ is a \emph{root information set}. For any pair of sequences $\sigma, \sigma' \in \Sigma_i^*$, with $\sigma = (j,a)$ and $\sigma' = (j', a')$, we write $\sigma \prec \sigma'$ if the sequence of actions encountered from the root of the tree to any node in $j'$ includes selecting action $a$ at some node from information set $j$. Further, by convention, we let $\emptyseq \prec \sigma$ for any $\sigma \in \Sigma_i^*$.

\paragraph{Sequence-form strategies} The strategy of a player specifies a probability distribution for every information set encountered in the tree. Assuming \emph{perfect recall}---players never forget acquired information---a strategy can be represented via the \emph{sequence-form strategy polytope} $\cQ_i \subseteq \R_{\geq 0}^{|\Sigma_i|}$, defined as 
\begin{equation*}
    \cQ_i \defeq \left\{ \vq_i \in \R_{\geq 0}^{|\Sigma_i|} : \vq_i[\emptyseq] = 1, \vq_i[\sigma_j] = \sum_{a \in \cA_j} \vq_i[(j,a)], \forall j \in \cJ_i \right\}.
\end{equation*}

Under the sequence-form representation, learning in extensive-form games can be cast in the framework of online linear optimization described earlier in \Cref{sec:regret}; we refer to, for example, the work of~\citet{Farina21:Simple}.

Further, we let $\Pi_i \defeq \cQ_i \cap \{0,1\}^{|\Sigma_i|}$ be the set of \emph{deterministic} sequence-form strategies. Analogously, one can define the sequence-form polytope $\cQ_j$ rooted at information set $j \in \cJ_i$, and $\Pi_j := \cQ_j \cap \{0, 1\}^{|\Sigma_j|}$. We also use $\| \cQ_i\|_1$ to denote the maximum $\ell_1$-norm of a vector $\vq_i \in \cQ_i$. Finally, we denote by $\depth_i$ the depth of $i$'s subtree.

\paragraph{Trigger deviations and EFCE}

To formalize the connection between EFCE and the framework of $\Phi$-regret, we introduce \emph{trigger deviation functions}.

\begin{definition}[\nakedcite{Farina21:Simple}]
    \label{def:trigdev}
    A trigger deviation function with respect to a trigger sequence $\hat{\sigma} = (j,a) \in \Sigma_i^*$ and a continuation strategy $\hat{\vec{\pi}}_i \in \Pi_j$ is any linear mapping $f : \R^{|\Sigma_i|} \to \R^{|\Sigma_i|}$ such that
    \begin{itemize}
        \item $f(\vec{\pi}_i) = \vec{\pi}_i$ for all $\vec{\pi}_i \in \Pi_i$ such that $\vec{\pi}_i[\hat{\sigma}] = 0$; \item Otherwise, for all $\vec{\pi}_i \in \Pi_i$,
        \begin{equation*}
            f(\vec{\pi}_i)[\sigma] =
            \begin{cases}
             \vec{\pi}_i[\sigma] \quad  \text{if } \sigma \not\succeq j, \\
             \hat{\vec{\pi}}_i[\sigma] \quad \text{if } \sigma \succeq j.
        \end{cases}
        \end{equation*}
    \end{itemize}
\end{definition}

We denote by $\Psi_i$ the convex hull of all trigger deviation functions---over all trigger sequences and deterministic continuation strategies; $\Psi_i$-regret is referred to as \emph{trigger regret}. In an \emph{extensive-form correlated equilibrium (EFCE)}~\citep{Stengel08:Extensive} no trigger deviation by any player can improve the utility of that player, leading to the following connection.

\begin{theorem}[\nakedcite{Farina21:Simple}]
    \label{prop:efce-trigger}
    If each player $i \in \range{n}$ incurs trigger regret $\reg_{\Psi_i}^T$ after $T$ repetitions of the game, the average product distribution of play is a $\frac{1}{T} \max_{i \in \range{n}} \reg^T_{\Psi_i}$-approximate EFCE.\looseness-1
\end{theorem}

Moreover, extensive-form \emph{coarse} correlated equilibria (EFCCE)~\citep{Farina20:Coarse} are defined analogously based on \emph{coarse trigger deviations} $\Tilde{\Psi}_i$; the difference is that in EFCCE the player decides whether to follow the recommendation \emph{before} actually seeing the recommendation at that information set (see \Cref{sec:addprels} for the definition and specific examples).
\section{Near-Optimal Learning for EFCE}
\label{sec:res}

In this section, we establish our main result: efficient learning dynamics with $O(\log T)$ per-player trigger regret; this is made precise in \Cref{cor:trigger}, the informal version of which was stated earlier in \Cref{theorem:efce}. 

This section is organized as follows: \Cref{sec:regtrig} analyzes the regret of the algorithm operating over the set of trigger deviation functions; \Cref{sec:fixedpoints} provides a refined characterization for the corresponding fixed points; and, finally, \Cref{sec:together} combines all the previous pieces to arrive at~\Cref{cor:trigger}. All the proofs from this section are deferred to~\Cref{sec:circuits}.

\subsection{Regret Minimizer for Trigger Deviations}
\label{sec:regtrig}

\paragraph{The algorithm} Our construction for minimizing trigger regret uses the template of~\citet{Gordon08:No} (\Cref{algo:Gordon}), and in particular, the approach of~\citet{Farina21:Simple} in order to construct an external regret minimizer for the set $\Psi_i$ (a similar approach also applies for the set of coarse trigger deviations $\Tilde{\Psi}_i$). More precisely, that construction leverages one separate regret minimizer $\regdep_{\hat{\sigma}}$ for every possible \emph{trigger sequence} $\hat{\sigma} \in \Sigma_i^*$ (recall \Cref{def:trigdev}). In particular, $\regdep_{\hat{\sigma}}$, with $\hat{\sigma} = (j,a)$, is---after performing an affine transformation---operating over sequence-form vectors $\vq_{\hat{\sigma}} \in \cQ_j$ (rooted at information set $j \in \cJ_i$). Then, those regret minimizers are combined using a regret minimizer $\regdep_{\triangle}$ operating over the simplex $\Delta(\Sigma_i^*)$. The first key ingredient in our construction is the use of a \emph{logarithmic regularizer}. More specifically, we instantiate each regret minimizer with \algoshort, a recent algorithm due to~\citet{Farina22:Near}. $\algoshort$ is an instance of~\eqref{eq:OFTRL}~\citep{Syrgkanis15:Fast} with logarithmic regularization; the main twist is that $\algoshort$ operates over an appropriately \emph{lifted space}. The overall construction is given in \Cref{algo:overall}.

\begin{algorithm}[!ht]
    \SetAlgoLined
    \SetInd{2.3mm}{2.3mm}
    \DontPrintSemicolon
    \KwData{\begin{itemize}[noitemsep,leftmargin=1.5mm]
        \item $\regdep_{\triangle} \leftarrow \algoshort(\eta_{\triangle})$ over $\Delta(\Sigma_i^*)$
        \item $\{ \regdep_{\hat{\sigma}} \leftarrow \algoshort(\eta)$ over $\cQ_j \}_{\hat{\sigma} = (j,\cdot) \in \Sigma_i^* }$
    \end{itemize}
    }    
        \Fn{$\nextstrat()$}{
            $\vec{\lambda}_i^{(t)} \leftarrow \regdep_{\triangle}.\nextstrat()$ \\
            $\vec{X}^{(t)}_{\hat{\sigma}} \leftarrow \regdep_{\hat{\sigma}}.\nextstrat()$, for all $\hat{\sigma} \in \Sigma_i^*$\\
            $\phi_i^{(t)} \leftarrow \sum_{\hat{\sigma} \in \Sigma_i^*} \vec{\lambda}_i^{(t)}[\hat{\sigma}] \vec{X}^{(t)}_{\hat{\sigma}}$ \label{line:convexcomb} \\
            $\vec{x}_i^{(t)} \leftarrow \fp(\phi_i^{(t)})$ \\
            \textbf{return} $\vec{x}_i^{(t)}$
        }
        \Hline{}
        \Fn{$\obsutil(\ut_i^{(t)})$}{
            Construct the utility $\vec{U}_i^{(t)} \leftarrow \ut_i^{(t)} \otimes\vec{x}^{(t)}_i$ \\
            $\regdep_{\hat{\sigma}}.\obsutil(\vec{U}_i^{(t)})$, for all $\hat{\sigma} \in \Sigma_i^*$ \\
            $\regdep_{\triangle}.\obsutil((\langle \vec{X}^{(t)}_{\hat{\sigma}}, \Vec{U}_i^{(t)} \rangle)_{\hat{\sigma} \in \Sigma_i^*})$
        }
    \caption{$\Psi_i$-Regret Minimizer}
    \label{algo:overall}
\end{algorithm}

For our purposes, we first apply~\citep[Proposition 2 and Corollary 1]{Farina22:Near} to obtain a suitable RVU bound for each regret minimizer $\regdep_{\hat{\sigma}}$ instantiated with \algoshort, for each $\hat{\sigma} \in \Sigma_i^*$.

\begin{restatable}{lemma}{rvusigma}
    \label{prop:rvu-sigma}
    Fix any $\hat{\sigma} \in \Sigma_i^*$, and let $\reg^T_{\hat{\sigma}}$ be the regret of $\regdep_{\hat{\sigma}}$ up to time $T \geq 2$. For any $\eta \leq \frac{1}{256\|\cQ_i\|_1}$, $\max\{0, \reg^T_{\hat{\sigma}}\}$ can be upper bounded by 
    \begin{align}
        \frac{2|\Sigma_i| \log T}{\eta} + 16 \eta \|\cQ_i\|_1^2 \sum_{t=1}^{T-1} \|\Ut_i^{(t+1)} - \Ut_i^{(t)}\|_\infty^2  - \frac{1}{512 \eta} \sum_{t=1}^{T-1} \| \vq^{(t+1)}_{\hat{\sigma}} - \vq^{(t)}_{\hat{\sigma}} \|^2_{\vq_{\hat{\sigma}}^{(t)}, \infty}.\label{eq:rvu-sigma}
    \end{align}
\end{restatable}
A few remarks are in order. First, we recall that $\Ut_i^{(t)} \defeq \ut_i^{(t)} \otimes \vx_i^{(t)}$, in accordance to~\Cref{algo:overall}. Also, $\eta > 0$ denotes the (time-invariant) learning rate of $\algoshort$. Furthermore, for $\hat{\sigma} = (j,a) \in \Sigma_i^*$, in \Cref{prop:rvu-sigma} we used the notation
\begin{equation*}
    \| \vq^{(t+1)}_{\hat{\sigma}} - \vq^{(t)}_{\hat{\sigma}} \|_{\vq_{\hat{\sigma}}^{(t)}, \infty} \defeq \max_{\sigma \in \Sigma_j} \left| 1 - \frac{\vq^{(t+1)}_{\hat{\sigma}}[\sigma]}{\vq^{(t)}_{\hat{\sigma}}[\sigma]} \right|.
\end{equation*}
\Cref{prop:rvu-sigma} establishes an RVU bound (\Cref{def:RVU}), but with two important refinements. First, the bound applies to $\max\{0, \reg^T_{\hat{\sigma}}\}$, instead of $\reg^T_{\hat{\sigma}}$, ensuring that~\eqref{eq:rvu-sigma} is nonnegative. Further, the local norm appearing in~\eqref{eq:rvu-sigma} will also be crucial for our argument in the sequel (\Cref{theorem:rational}). 

Next, similarly to \Cref{prop:rvu-sigma}, we obtain a regret bound for $\regdep_{\triangle}$, the regret minimizer ``mixing'' over all $\{ \regdep_{\hat{\sigma}}\}_{\hat{\sigma} \in \Sigma_i^*}$.

\begin{restatable}{lemma}{rvutri}
    \label{prop:rvu-tri}
    Let $\reg^T_{\triangle}$ be the regret of $\regdep_{\triangle}$ up to time $T \geq 2$. For any $\eta_{\triangle} \leq \frac{1}{512 |\Sigma_i|}$, $\max \{0, \reg^T_{\triangle} \}$ can be upper bounded by
    \begin{align*}
    \frac{2|\Sigma_i| \log T}{\etatri} +16\etatri |\Sigma_i|^2 \sum_{t=1}^{T-1} \|\uttri^{(t+1)} - \uttri^{(t)}\|^2_{\infty} -\frac{1}{512\etatri} \sum_{t=1}^{T-1} \|\vlam_i^{(t+1)} - \vlam_i^{(t)} \|^2_{\vlam_i^{(t)}, \infty}. 
    \end{align*}
\end{restatable}

Here, we used the notation $\uttri^{(t)}[\hat{\sigma}] \defeq \langle \vX_{\hat{\sigma}}^{(t)}, \Ut_i^{(t)} \rangle $, where $\vX^{(t)}_{\hat{\sigma}}$ is the output of $\regdep_{\hat{\sigma}}$, for each $\hat{\sigma} \in \Sigma_i^*$; that is, $\vX_{\hat{\sigma}}^{(t)} \in \R^{|\Sigma_i| \times |\Sigma_i|}$ transforms sequence-form vectors based on the continuation strategy $\vq^{(t)}_{\hat{\sigma}}$ below the trigger sequence $\hat{\sigma}$ (recall \Cref{def:trigdev}). We also note that $\vec{\lambda}_i^{(t)} \in \Delta(\Sigma^*_i)$ above represents the output of $\regdep_{\triangle}$ at time $t$.

We next use \Cref{theorem:gordon} to obtain a bound for $\reg^T_{\Psi_i}$, the $\Psi_i$-regret of the overall construction (\Cref{algo:overall}).

\begin{restatable}{proposition}{hull}
    \label{prop:rvu-hull}
    For any $T \in \N$,
    \begin{equation*}
        \max\{0, \reg^T_{\Psi_i} \} \leq \max\{0, \reg_{\triangle}^T\} + \sum_{\hat{\sigma} \in \Sigma_i^*} \max\{0, \reg^T_{\hat{\sigma}}\}.
    \end{equation*}
\end{restatable}

This uses the \textit{regret circuit} for the convex hull~\citep{Farina19:Regret} to combine all the regret minimizers $\{ \regdep_{\hat{\sigma}}\}_{\hat{\sigma} \in \Sigma_i^*}$ via $\regdep_{\triangle}$ into an external regret minimizer for the set $\Psi_i$; by virtue of \Cref{theorem:gordon}, the external regret of the induced algorithm is equal to the $\Psi_i$-regret ($\reg^T_{\Psi_i}$) of player $i$. There is, however, one crucial twist: the guarantee of~\citet{Farina19:Regret} would give a bound in terms of $\max_{\hat{\sigma} \in \Sigma_i^*} \reg^T_{\hat{\sigma}}$, instead of $\sum_{\hat{\sigma} \in \Sigma_i^*} \reg^T_{\hat{\sigma}}$; this is problematic for obtaining near-optimal rates as \emph{it breaks the RVU property over the convex hull}. In general, it is not clear how to bound the maximum of the regrets by their sum since (external) regret \emph{can be negative}. This is, in fact, a recurrent obstacle encountered in this line of work~\citep{Syrgkanis15:Fast}, and it is precisely the reason why approaches based on regret decomposition---in the spirit of CFR~\citep{Zinkevich07:Regret}---failed to bring rates better than $T^{-3/4}$~\citep{Farina19:Stable}. \Cref{prop:rvu-hull} circumvents those obstacles by establishing bounds in terms of \emph{nonnegative measures of regret}. 

\begin{remark}[Near-optimal regret via CFR-type algorithms]
    An important byproduct of our techniques, and in particular \Cref{prop:rvu-hull} along with RVU bounds for nonnegative measures of regret~\citep{Anagnostides22:Uncoupled}, is the first near-optimal $O(\log T)$ regret bound for CFR-type algorithms in general games, a question that has been open even in (two-player) zero-sum games (see the discussion by~\citet{Farina19:Stable}).\footnote{\citet{Liu22:The} very recently obtained near-optimal rates in zero-sum games, though with very different techniques.}
\end{remark}

\subsection{Characterizing the Fixed Points}
\label{sec:fixedpoints}

Next, our main goal is to obtain an RVU bound for $\max\{0, \reg^T_{\Psi_i}\}$, but cast in terms of the player's strategies $(\vx_i^{(t)})_{1 \leq t \leq T}$, as well as the utilities $(\ut_i^{(t)})_{1 \leq t \leq T}$ observed by that player. In particular, in light of \Cref{prop:rvu-sigma,prop:rvu-tri}, the crux is to appropriately bound $\|\vlam_i^{(t+1)} - \vlam_i^{(t)} \|_{\vlam_i^{(t)}, \infty}$ and $ \sum_{\hat{\sigma} \in \Sigma_i^*} \| \vq^{(t+1)}_{\hat{\sigma}} - \vq^{(t)}_{\hat{\sigma}} \|_{\vq_{\hat{\sigma}}^{(t)},\infty}$ in terms of $\|\vx_i^{(t+1)} - \vx_i^{(t)}\|$---the deviation of the player's strategy at every time $t$. To do so, we prove the following key result.

\begin{restatable}{lemma}{gamma}
    \label{theorem:rational}
    Let $\vX_i^{(t)} \in \R_{> 0}^{D}$ be defined for every time $t \in \N$, for some $D \in \N$. Further, suppose that for every time $t \in \N$ and $\sigma \in \Sigma_i$,
    \begin{equation}
        \label{eq:rational}
        \vx_i^{(t)} [\sigma] = \sum_{k=1}^m \frac{p_{\sigma, k}(\vX_i^{(t)})}{q_{\sigma, k}(\vX_i^{(t)})},
    \end{equation}
    for some multivariate polynomials $\{p_{\sigma, k}\}, \{q_{\sigma, k}\}$ with positive coefficients and maximum degree $\deg_i \in \N$. If
    \begin{equation}
        \label{eq:mul-stab-X}
        \max_{\edg \in \range{D}} \left| 1 - \frac{\vX_i^{(t+1)}[\edg]}{\vX_i^{(t)}[\edg]} \right| \leq \frac{100}{256 \deg_i},
    \end{equation}
    it holds that
    \begin{equation*}
    \| \vx_i^{(t+1)} - \vx_i^{(t)} \|_1 \leq (4 \| \cQ_i\|_1 \deg_i) \max_{\edg \in \range{D}} \left| 1 - \frac{\vX_i^{(t+1)}[\edg]}{\vX_i^{(t)}[\edg]} \right|.
\end{equation*}
\end{restatable}

We recall that, based on \Cref{algo:Gordon}, the final strategy $\vx_i^{(t)}$ is simply a fixed point of $\phi_i^{(t)} \in \Psi_i^{(t)}$, where $\phi_i^{(t)}$ is a function of $\vX_i^{(t)} = ( \vlam_i^{(t)}, (\vq^{(t)}_{\hat{\sigma}})_{\hat{\sigma} \in \Sigma_i^*})$ (\Cref{line:convexcomb}). \Cref{eq:rational} postulates that the fixed point is given by a rational function with positive coefficients. Taking a step back, let us clarify that assumption in the context of the no-swap-regret algorithm of~\citet{Blum07:From}, a specific instance of \Cref{algo:Gordon}. In that algorithm, the fixed point is a stationary distribution of the underlying stochastic matrix $\vX_i$; hence, \eqref{eq:rational} is simply a consequence of the \emph{Markov chain tree theorem} (see \citep{Anantharam89:A}), with degree in the order of the rank of the corresponding stochastic matrix.

While insisting on having positive coefficients in \Cref{theorem:rational} may seem restrictive at first glance, in \Cref{prop:poscoeffs,prop:poscoeffs-tree} (in \Cref{sec:circuits}) we show that, in fact, it comes without any loss under sequence-form vectors (modulo an additive constant term). We further remark that the degree of the rational function is a measure of the complexity of the fixed points, as it will be highlighted in \Cref{prop:fp-efcce,prop:fp-efce} below. Finally, the property in \eqref{eq:mul-stab-X} will be satisfied for our construction since the regret minimizers we employ guarantee \emph{multiplicative stability} (as we formally show in \Cref{lemma:q-mul-stab,lemma:l-mul-sta}), meaning that the ratio of any two consecutive coordinates is close to $1$; this refined notion of stability is ensured by the use of the logarithmic regularizer.\looseness-1

We now establish that assumption~\eqref{eq:rational} is satisfied for transformations in $\Psi_i$ with only a moderate degree. First, as a warm-up, we consider fixed points associated with \emph{coarse} trigger deviation functions $\Tilde{\Psi}_i$.

\begin{restatable}{proposition}{efccerational}
    \label{prop:fp-efcce}
    Let $\phi_i^{(t)} \in \Tilde{\Psi}_i$ be a transformation defined by $\vX_i^{(t)} = ( \vlam_i^{(t)}, (\vq^{(t)}_{j})_{j \in \cJ_i}) \in \R^D_{>0} $, for some $D \in \N$ and time $t \in \N$. The unique fixed point $\vx_i^{(t)}$ of $\phi_i^{(t)}$ satisfies \eqref{eq:rational} with $\deg_i \leq 2 \depth_i$.
\end{restatable}

This property is established by leveraging the closed-form characterization for the fixed points associated with EFCCE given by~\citet{Anagnostides22:Faster}. Next, let us focus on the fixed points of trigger deviation functions. Unlike EFCCE, determining such fixed points requires computing stationary distributions of Markov chains along paths of the tree, commencing from the root and gradually making way towards the leaves~\citep{Farina21:Simple}; this substantially complicates the analysis. Nevertheless, we leverage a refined characterization of the stationary distribution at every information set~\citep{Anagnostides22:Faster} to obtain the following.

\begin{restatable}{proposition}{rationalefce}
    \label{prop:fp-efce}
    Let $\phi_i^{(t)} \in \Psi_i$ be a transformation defined by $\vX_i^{(t)} = ( \vlam_i^{(t)}, (\vq^{(t)}_{\hat{\sigma}})_{\hat{\sigma} \in \Sigma^*_i}) \in \R^D_{>0} $, for some $D \in \N$ and time $t \in \N$. The (unique) fixed point $\vx_i^{(t)}$ of $\phi_i^{(t)}$ satisfies \eqref{eq:rational} with $\deg_i \leq 2 \depth_i |\cA_i|$, where $|\cA_i| \defeq \max_{j \in \cJ_i} |\cA_j|$.
\end{restatable}

In proof, we show that augmenting a ``partial fixed point'' at a new (successor) information set can only increase the degree of the rational function by an additive factor of $2 |\cA_i|$; \Cref{prop:fp-efce} then follows by induction. It is crucial to note that using the Markov chain tree theorem directly at every information set would only give a bound on the degree that could be exponential in the description of the extensive-form game. Next, we combine \Cref{prop:fp-efce} with \Cref{theorem:rational} to derive the following key inequality.

\begin{restatable}{lemma}{gammaefce}
    \label{lemma:gamma-efce}
    Consider any parameters $\eta \leq \frac{1}{256 \|\cQ_i\|_1 \deg_i }$ and $\etatri \leq \frac{1}{512|\Sigma_i| \deg_i}$, where $\deg_i \defeq 2 |\cA_i| \depth_i$. Then, for any time $t \in \range{T-1}$,
    \begin{align*}
        \|\vx_i^{(t+1)} - \vx_i^{(t)}\|_1 \leq 8 \|\cQ_i\|_1 |\cA_i| \depth_i M(\vX_i^{(t)}), 
    \end{align*}
    where $M(\vX_i^{(t)})$ is defined as 
    \begin{equation*}
        \max\left\{ \max_{\hat{\sigma} \in \Sigma_i^*} \left| 1 - \frac{\vlam_i^{(t+1)}[\hat{\sigma}]}{\vlam_i^{(t)}[\hat{\sigma}]}\right|, \max_{\hat{\sigma}  \in \Sigma_i^*} \max_{\sigma \in \Sigma_j} \left| 1 - \frac{\vq^{(t+1)}_{\hat{\sigma}}[\sigma]}{\vq^{(t)}_{\hat{\sigma}}[\sigma]} \right| \right\}.
    \end{equation*}
\end{restatable}

\subsection{Putting Everything Together}
\label{sec:together}

We now combine \Cref{lemma:gamma-efce} with \Cref{prop:rvu-sigma,prop:rvu-tri,prop:rvu-hull}, as well as some further manipulations of the utilities $(\Ut_i^{(t)})_{1 \leq t \leq T}$ (\Cref{prop:rvu-sigma}) and $(\uttri^{(t)})_{1 \leq t \leq T}$ (\Cref{prop:rvu-tri}) to derive the following RVU bound.

\begin{restatable}{corollary}{corrvu}
    \label{cor:rvu}
    Suppose that $\eta \leq \frac{1}{2^{12} |\Sigma_i|^{1.5} \|\cQ_i\|_1 \deg_i }$ and $\etatri = \frac{1}{2|\Sigma_i|}\eta$, where $\deg_i \defeq 2 |\cA_i| \depth_i$. For any $T \geq 2$, $\max\{0, \reg^T_{\Psi_i}\}$ can be upper bounded by
    \begin{align*}
        \frac{8|\Sigma_i|^2 \log T}{\eta} + 256 \eta |\Sigma_i|^3 \sum_{t=1}^{T-1} \|\ut_i^{(t+1)} - \ut_i^{(t)}\|_\infty^2 - \frac{1}{2^{15} \eta \deg_i^2 \|\cQ_i\|_1^2} \sum_{t=1}^{T-1} \|\vx_i^{(t+1)} - \vx_i^{(t)}\|_1^2.
    \end{align*}
\end{restatable}

Given that the RVU bound in \Cref{cor:rvu} has been obtained for $\max\{0, \reg^T_{\Psi_i}\}$, a nonnegative measure of regret, we can show that the second-order path length of the dynamics when all players follow \Cref{algo:overall} is bounded by $O(\log T)$; that is, 
$$\sum_{t=1}^{T-1} \sum_{i=1}^n \|\vx_i^{(t+1)} - \vx_i^{(t)}\|_1^2 = O(\log T).$$ 
This step is formalized in \Cref{cor:pathlength} (in \Cref{sec:circuits}), and follows the technique of~\citet{Anagnostides22:Uncoupled}, leading to our main result; below we use the notation $|\Sigma| \defeq \max_{i \in \range{n}} |\Sigma_i|$, and similarly for the other symbols (namely, $\|\cQ\|_1, |\cA|$ and $\depth$).

\begin{theorem}
    \label{cor:trigger}
    If all players employ \Cref{algo:overall}, the trigger regret of each player $i \in \range{n}$ after $T$ repetitions will be bounded as
    \begin{equation}
        \label{eq:ub}
        \reg^T_{\Psi_i} \leq C n |\Sigma|^{3.5} \|\cQ\|_1 |\cZ| |\cA| \depth \log T,
    \end{equation}
    for a universal constant $C > 0$.
\end{theorem}

While the above theorem applies when \emph{all} players follow the prescribed protocol, it is easy to ensure at the same time that the trigger regret of each player will grow as $O(\sqrt{T})$ even if the rest of the players are instead acting so as to minimize that player's utility~\citep{Anagnostides22:Uncoupled}.

As we highlight in~\Cref{section:conclusions}, improving the dependence of~\eqref{eq:ub} on the underlying parameters of the game is an important direction for future work. For EFCCE, in accordance to \Cref{prop:fp-efcce}, we obtain a slightly improved regret bound (see \Cref{cor:efcce} in \Cref{sec:circuits}).

\begin{remark}[Beyond EFCE]
    \label{remark:beyond}
    While we have focused primarily on obtaining near-optimal guarantees for trigger regret, corresponding to EFCE, our techniques apply more broadly to $\Phi$-regret minimization under two conditions: i) the fixed point of any $\phi \in \Phi$ should admit a characterization as a rational function per~\eqref{eq:rational}; and (ii) one can efficiently perform projections to the set $\Phi$ (in the sense of~\citet{Farina22:Near}). For example, we believe that those two conditions are met even when the set of deviations $\Phi$ includes all possible linear transformations, which is of course stronger than trigger regret.
\end{remark}
\section{Experimental Results}
\label{section:experiments}

Finally, in this section we experimentally verify our theoretical results on several common benchmark extensive-form games: (i) $3$-player \emph{Kuhn poker}~\citep{Kuhn53:Extensive}; (ii) $2$-player \emph{Goofspiel}~\citep{Ross71:Goofspiel}; and (iii) $2$-player \emph{Sheriff}~\citep{Farina19:Correlation}. We note that none of the above is a two-player zero-sum game. A detailed description of the game instances we use is included in \Cref{appendix:games}.

In accordance to \Cref{cor:trigger}, we instantiate each local regret minimizer using \algoshort, and all players use the same learning algorithm. For simplicity we use the same learning rate $\eta > 0$ for all the local regret minimizers, which is treated as a hyperparameter in order to obtain better empirical performance. In particular, after a very mild tuning process, we chose $\eta = 1$ for all our experiments. We compare the performance of our algorithm with that of two other popular regret minimizers: 1) CFR with regret matching (RM)~\citep{Zinkevich07:Regret}, meaning that every local regret minimizer $\regdep_{\hat{\sigma}}$ uses CFR (with RM) and $\regdep_{\triangle}$ (which is an algorithm for the simplex) also uses RM; and 2) CFR$^+$ with RM$^+$~\citep{Tammelin15:Solving}. We did not employ alternation or linear averaging, two popular tricks that accelerate convergence in zero-sum games,
as it is not known if those techniques retain convergence in our setting.

\begin{figure*}[!ht]
    \centering
    \resizebox{.98\textwidth}{!}{\input{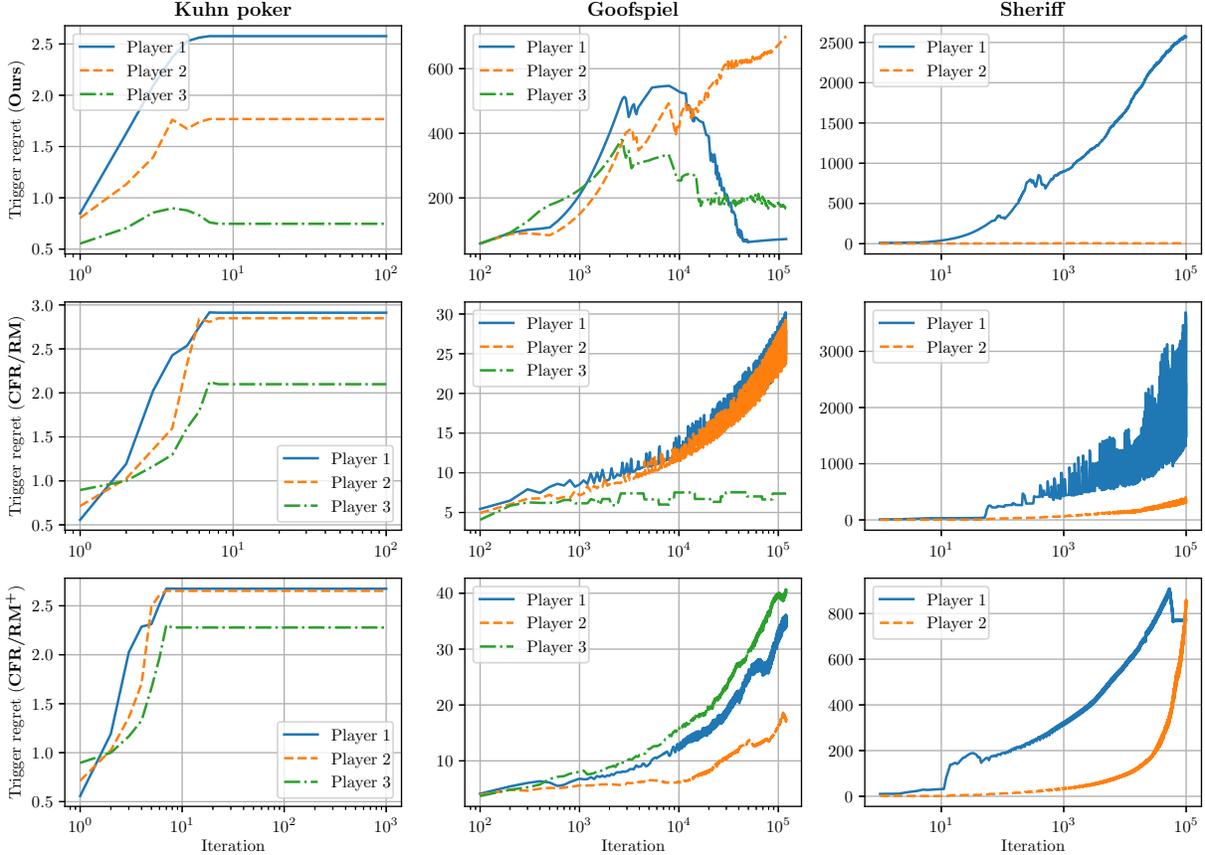}}
    \caption{Trigger regret of each player on (i) Kuhn poker (left); (ii) Goofspiel (center); and (iii) Sheriff (right). Every row corresponds to a different algorithm, starting from ours in the first one. The $x$-axis indicates the iteration, while the $y$-axis indicates the corresponding trigger regret for each player. We emphasize that the $x$-axis is logarithmic.}\vspace{-1mm}
    \label{fig:plot}
\end{figure*}

Our findings are illustrated in \Cref{fig:plot}. As predicted by our theory (\Cref{cor:trigger}), the trigger regret of all players appears to grow as $O(\log T)$ (the $x$-axis is logarithmic), implying convergence to the set of EFCE with a rate of $\frac{\log T}{T}$. In contrast, although the trigger regret experienced by the other regret minimizers is sometimes smaller compared to our algorithm, their asymptotic growth apparently exhibits an unfavorable exponential increase, meaning that their trigger regret grows as $\omega(\log T)$, with the exception of $3$-player Kuhn poker. In fact, for Kuhn poker we see that the learning dynamics actually converge to a Nash equilibrium after only a few iterations, but this is not a typical behavior beyond two-player zero-sum games. Indeed, for the other two games in \Cref{fig:plot} we do not have convergence to a Nash equilibrium, although---as predicted by our theory---we observe convergence to EFCE (since the players' \emph{average} trigger regrets approach to $0$). The overall conclusion is that our algorithm should be preferred for a sufficiently high precision.

It is also worth pointing out that we obtained qualitatively similar regret bounds for \emph{coarse} trigger regret---associated with EFCCE. 

\section{Conclusions and Future Research}
\label{section:conclusions}

In this paper, we established the first near-optimal $\frac{\log T}{T}$ rates of convergence to extensive-form correlated equilibria, thereby extending recent work from normal-form games to the substantially more complex class of imperfect-information extensive-form games. Our approach for obtaining near-optimal $\Phi$-regret guarantees can be in fact further extended even beyond extensive-form games, as long as the fixed points admit the characterization imposed by \Cref{theorem:rational}. Our techniques also have an independent interest in deriving near-optimal rates using the regret-decomposition approach, a question that has previously remained elusive even in two-player zero-sum games~\citep{Farina19:Stable}. In particular, we initiated the study of regret circuits---in the sense of~\citet{Farina19:Regret}---that \emph{preserve the RVU property}, and we established a new composition result for the convex hull, which has been the main obstacle in prior approaches~\citep{Farina19:Stable}.

There are still many interesting avenues for future research related to our work. While our trigger-regret bounds are near-optimal in terms of the dependence on $T$ (\Cref{cor:trigger}), the dependence on the parameters of the game in~\eqref{eq:ub} can likely be improved. Establishing near-optimal trigger regret under dynamics that do not employ logarithmic regularization, such as optimistic hedge, could be a helpful step in that direction, but that is currently a challenging open problem; it is plausible that the techniques of~\citet{Anagnostides21:Near} in conjunction with the regret bounds of~\citet{Daskalakis21:Near} could be useful in that direction, although the combinatorial complexity of trigger deviation functions poses considerable challenges. 

Another interesting problem is to characterize the set of transformations $\Phi$---beyond trigger regret---under which our techniques are applicable (see \Cref{remark:beyond}). In a similar vein, we suspect that our approach leads to near-optimal convergence to the set of \emph{behavioral correlated equilibria (BCE)}, which corresponds to minimizing swap regret at every information set locally~\citep{Morrill21:Hindsight}.

\section*{Acknowledgements}

We are grateful to the anonymous reviewers at ICML 2023 for a number of helpful suggestions and corrections. The work of Prof. Sandholm's research group is funded by the National Science Foundation under grants IIS1901403, CCF-1733556, and the ARO under award W911NF2210266.

\bibliography{main}

\iftrue
\newpage
\appendix
\clearpage

\section{Additional Preliminaries}
\label{sec:addprels}

In this section, we provide some additional background on extensive-form games and (coarse) trigger deviation functions.

\paragraph{An illustrative example} First, to clarify some of the concepts we introduced earlier in \Cref{section:prel}, we consider the simple two-player EFG of \Cref{fig:example}. White round nodes correspond to player $1$, while black round nodes to player $2$. We use square nodes to represent terminal nodes (or leaves). As illustrated in \Cref{fig:example}, player $1$ has two information sets, denoted by $\cJ_1 \defeq \{\Large{\textsc{a}}, \Large{\textsc{b}}\}$, each containing two nodes. Further, the set of sequences of player $1$ can be represented as $\Sigma_1 \defeq \{\emptyseq, \seq{1}, \seq{2}, \seq{3}, \seq{4}\}$; here, we omitted specifying the corresponding information set since we use different symbols for actions belonging to different information sets.

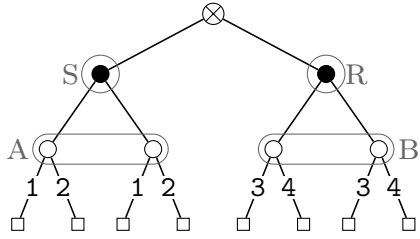
\begin{figure}[H]
    \centering
    \scalebox{1}{\begin{tikzpicture}[baseline=0cm]
       \def\dcha{1.5}
       \def\done{.7}
       \def\dtwo{.4}
       \node[chanode] (A) at (0, -.2) {};
       \node[pl1node] (B) at (-\dcha,-1) {};
       \node[pl1node] (C) at (\dcha,-1) {};
       \node[pl2node] (D) at (-\dcha-\done,-2) {};
       \node[pl2node] (E) at (-\dcha+\done,-2) {};
       \node[pl2node] (F) at (\dcha-\done,-2) {};
       \node[pl2node] (G) at (\dcha+\done,-2) {};
       \node[termina] (l1) at (-\dcha-\done-\dtwo,-3) {};
       \node[termina] (l2) at (-\dcha-\done+\dtwo,-3) {};
       \node[termina] (l3) at (-\dcha+\done-\dtwo,-3) {};
       \node[termina] (l4) at (-\dcha+\done+\dtwo,-3) {};
       \node[termina] (l5) at (\dcha-\done-\dtwo,-3) {};
       \node[termina] (l6) at (\dcha-\done+\dtwo,-3) {};
       \node[termina] (l7) at (\dcha+\done- \dtwo,-3) {};
       \node[termina] (l8) at (\dcha+\done+\dtwo,-3) {};

       \draw[semithick] (A) -- (B)
                  --node {} (D)
                  --node[fill=white,inner sep=.9] {\seq{1}} (l1);
       \draw[semithick] (B) --node {} (E)
                  --node[fill=white,inner sep=.9] {\seq{1}} (l3);
       \draw[semithick] (D) --node[fill=white,inner sep=.9] {\seq{2}} (l2);
       \draw[semithick] (E) --node[fill=white,inner sep=.9] {\seq{2}} (l4);
       \draw[semithick] (A) -- (C)
                  --node {} (F)
                  --node[fill=white,inner sep=.9] {\seq{3}} (l5);
       \draw[semithick] (C) --node {} (G)
                  --node[fill=white,inner sep=.9] {\seq{3}} (l7);
       \draw[semithick] (F) --node[fill=white,inner sep=.9] {\seq{4}} (l6);
       \draw[semithick] (G) --node[fill=white,inner sep=.9] {\seq{4}} (l8);

       \draw[black!60!white] (B) circle (.25);
       \node[black!60!white]  at ($(B) + (-.4, 0)$) {\Large\textsc{s}};

       \draw[black!60!white] (C) circle (.25);
       \node[black!60!white]  at ($(C) + (.4, 0)$) {\Large\textsc{r}};

       \draw[black!60!white] ($(D) + (0, .2)$) arc (90:270:.2);
       \draw[black!60!white] ($(D) + (0, .2)$) -- ($(E) + (0, .2)$);
       \draw[black!60!white] ($(D) + (0, -.2)$) -- ($(E) + (0, -.2)$);
       \draw[black!60!white] ($(E) + (0, -.2)$) arc (-90:90:.2);
       \node[black!60!white]  at ($(D) + (-.4, 0)$) {\Large\textsc{a}};

       \draw[black!60!white] ($(F) + (0, .2)$) arc (90:270:.2);
       \draw[black!60!white] ($(F) + (0, .2)$) -- ($(G) + (0, .2)$);
       \draw[black!60!white] ($(F) + (0, -.2)$) -- ($(G) + (0, -.2)$);
       \draw[black!60!white] ($(G) + (0, -.2)$) arc (-90:90:.2);
       \node[black!60!white]  at ($(G) + (.4, 0)$) {\Large\textsc{b}};
\end{tikzpicture}}  
    \caption{Example of a two-player EFG.}
    \label{fig:example}
\end{figure}

\paragraph{Trigger deviation functions} It will be convenient to represent trigger deviation functions, in the sense of \Cref{def:trigdev}, as follows.

\begin{definition}[\nakedcite{Farina21:Simple}]
    \label{def:matM}
    Let $\hat{\sigma} = (j,a) \in \Sigma_i^*$, and $\vq \in \cQ_j$. We let $\mat{M}_{\hat{\sigma} \to \vq} \in \R^{|\Sigma_i| \times |\Sigma_i|}$ be a matrix, so that for any $\sigma_r, \sigma_c \in \Sigma_i$,
    \begin{equation*}
        \mat{M}_{\hat{\sigma} \to \vq} =
        \begin{cases}
            1 & \text{if } \sigma_c \not\succeq \hat{\sigma} \text{ and } \sigma_r = \sigma_c;\\
            \vq[\sigma_r] & \text{if } \sigma_c = \hat{\sigma} \text{ and } \sigma_r \succeq j; \text{ and} \\
            0 & \text{otherwise}.
        \end{cases}
    \end{equation*}
\end{definition}

We will let $\phi_{\hat{\sigma} \to \vq}$ denote the linear function $\vx \mapsto \mat{M}_{\hat{\sigma} \to \vq} \vx$, for some $\vq \in \cQ_j$. It is immediate to verify that for any $\hat{\sigma} =(j,a) \in \Sigma_i^*$ and $\vq \in \cQ_j$, $\phi_{\hat{\sigma} \to \vq}$ is a trigger deviation function in the sense of \Cref{def:trigdev}.  

To clarify \Cref{def:matM}, below we give two examples for the EFG of \Cref{fig:example}. If $\vq = (\frac{1}{2}, \frac{1}{2}) \in \Delta^2$, then

\[
\mat{M}_{\seq{1} \to \vq} =
\begin{blockarray}{cccccc}
& \emptyseq & \seq{1} & \seq{2} & \seq{3} & \seq{4} \\
\begin{block}{c (ccccc)}
  \emptyseq & 1 & 0 & 0 & 0 & 0 \\
  \seq{1} & 0 & \textbf{1/2} & 0 & 0 & 0 \\
  \seq{2} & 0 & \textbf{1/2} & 1 & 0 & 0 \\
  \seq{3} & 0 & 0 & 0 & 1 & 0 \\
  \seq{4} & 0 & 0 & 0 & 0 & 1 \\
\end{block}
\end{blockarray}, \quad
\mat{M}_{\seq{3} \to \vq} =
\begin{blockarray}{cccccc}
& \emptyseq & \seq{1} & \seq{2} & \seq{3} & \seq{4} \\
\begin{block}{c (ccccc)}
  \emptyseq & 1 & 0 & 0 & 0 & 0 \\
  \seq{1} & 0 & 1 & 0 & 0 & 0 \\
  \seq{2} & 0 & 0 & 1 & 0 & 0 \\
  \seq{3} & 0 & 0 & 0 & \textbf{1/2} & 0 \\
  \seq{4} & 0 & 0 & 0 & \textbf{1/2} & 1 \\
\end{block}
\end{blockarray}.
 \]

The following characterization can be readily extracted from~\citep{Farina21:Simple}.

\begin{claim}
    \label{claim:equiv-efce}
    Every transformation $\phi_i \in \Psi_i$ can be expressed as $\sum_{\hat{\sigma} \in \Sigma_i^*} \vlam_i[\hat{\sigma}] \phi_{\hat{\sigma} \to \vq_{\hat{\sigma}}}$, where $\vlam_i \in \Delta(\Sigma_i^*)$ and $\vq_{\hat{\sigma}} \in \cQ_j$ for $\hat{\sigma} = (j,a) \in \Sigma_i^*$.
\end{claim}
\paragraph{Coarse trigger deviation functions} Analogously, \emph{coarse} trigger deviation functions can be represented as follows.

\begin{definition}[\nakedcite{Anagnostides22:Faster}]
    \label{def:matM'}
    Let $j \in \cJ_i$ and $\vq \in \cQ_{j}$. We let $\mat{M}_{j \to \vq} \in \R^{|\Sigma_i| \times |\Sigma_i|}$ be a matrix, so that for any $\sigma_r, \sigma_c \in \Sigma_i$,
    \begin{equation*}
        \mat{M}_{j \to \vq} =
        \begin{cases}
            1 & \text{if } \sigma_c \not\succeq j \text{ and } \sigma_r = \sigma_c; \\
            \vq[\sigma_r] & \text{if } \sigma_c = \sigma_j \text{ and } \sigma_r \succeq j; \text{ and} \\
            0 & \text{otherwise}.
        \end{cases}
    \end{equation*}
\end{definition}

Unlike trigger deviations, which are ``triggered'' by a sequence, we point out that \emph{coarse} trigger deviations are ``triggered'' by an information set; see the work of~\citet{Farina20:Coarse} for a more detailed discussion on this point.

Returning to the example of \Cref{fig:example}, and letting again $\vq = (\frac{1}{2}, \frac{1}{2}) \in \Delta^2$,
\[
\mat{M}_{\Large{\textsc{A}} \to \vq} =
\begin{blockarray}{cccccc}
& \emptyseq & \seq{1} & \seq{2} & \seq{3} & \seq{4} \\
\begin{block}{c (ccccc)}
  \emptyseq & 1 & 0 & 0 & 0 & 0 \\
  \seq{1} & \textbf{1/2} & 0 & 0 & 0 & 0 \\
  \seq{2} & \textbf{1/2} & 0 & 0 & 0 & 0 \\
  \seq{3} & 0 & 0 & 0 & 1 & 0 \\
  \seq{4} & 0 & 0 & 0 & 0 & 1 \\
\end{block}
\end{blockarray}, \quad
\mat{M}_{\Large{\textsc{C}} \to \vq} =
\begin{blockarray}{cccccc}
& \emptyseq & \seq{1} & \seq{2} & \seq{3} & \seq{4} \\
\begin{block}{c (ccccc)}
  \emptyseq & 1 & 0 & 0 & 0 & 0 \\
  \seq{1} & 0 & 1 & 0 & 0 & 0 \\
  \seq{2} & 0 & 0 & 1 & 0 & 0 \\
  \seq{3} & \textbf{1/2} & 0 & 0 & 0 & 0 \\
  \seq{4} & \textbf{1/2} & 0 & 0 & 0 & 0 \\
\end{block}
\end{blockarray}.
 \]

Analogously to \Cref{claim:equiv-efce}, one can show the following characterization.

\begin{claim}
    \label{claim:equiv-efcce}
    Every transformation $\phi_i \in \Tilde{\Psi}_i$ can be expressed as $\sum_{j \in \cJ_i} \vlam_i[j] \phi_{j \to \vq_{j}}$, where $\vlam_i \in \Delta(\cJ_i)$ and $\vq_{j} \in \cQ_j$.
\end{claim}

The connection between coarse trigger deviation functions and EFCCE is illuminated in the following fact.

\begin{theorem}[\nakedcite{Anagnostides22:Faster}]
    \label{prop:efcce-trigger}
    If each player $i \in \range{n}$ incurs coarse trigger regret $\reg_{\Tilde{\Psi}_i}^T$ after $T$ repetitions of the game, the average product distribution of play is a $\frac{1}{T} \max_{i \in \range{n}} \reg^T_{\Tilde{\Psi}_i}$-approximate EFCCE.
\end{theorem}

\section{Omitted Proofs}
\label{sec:circuits}

In this section, we provide all the omitted proofs from the main body (\Cref{sec:res}). For the convenience of the reader, we restate each claim before proceeding with its proof.

\subsection{RVU Bounds for the Set of Deviations}

Let us fix a player $i \in \range{n}$. First, we prove \Cref{prop:rvu-sigma}. To this end, let us provide some auxiliary claims. Recall that, for each $\hat{\sigma} = (j,a) \in \Sigma_i^*$, $\regdep_{\hat{\sigma}}$ receives at every time $t$ the utility $\Ut_i^{(t)} \defeq \ut_i^{(t)} \otimes \vx_i^{(t)}$, and the next strategy is computed via~\algoshort~\citep{Farina22:Near}; namely, we first compute
$\Tilde{\vq}_{\hat
\sigma}^{(t)} \defeq (\Tilde{\vq}_{\hat
\sigma}^{(t)}[0], (\Tilde{\vq}_{\hat
\sigma}^{(t)}[e])_{e \in \Sigma_j} ) = (\lambda_{\hat{\sigma}}^{(t)}, \vy_{\hat{\sigma}}^{(t)}) \in \Tilde{\cQ_j}$, for a time $t \in \N$, as
\begin{equation}
    \label{eq:lrl}
    \argmax_{\Tilde{\vq}_{\hat{\sigma}} \in \Tilde{\cQ_j}} \left\{ \eta \left\langle \vS_{\hat{\sigma}}^{(t-1)}, \Tilde{\vq}_{\hat{\sigma}} \right\rangle + \sum_{\edg \in \Sigma_j \cup \{0\}}\log \Tilde{\vq}_{\hat{\sigma}}[\edg] \right\},
\end{equation}
where,
\begin{itemize}
    \item[(i)] $\Tilde{\cQ_j} \defeq \{ (\lambda_{\hat{\sigma}}, \vy_{\hat{\sigma}}) : \lambda_{\hat{\sigma}} \in [0,1], \vy_{\hat{\sigma}} \in \lambda_{\hat{\sigma}} \cQ_j \}$;
    \item[(ii)] $\Tilde{\Ut}_{\hat{\sigma}}^{(t)} \defeq (- \langle  \vq_{\hat{\sigma}}^{(t)}, \Ut_{\hat{\sigma}}^{(t)} \rangle, \Ut_{\hat{\sigma}}^{(t)})$, where in turn $\Ut^{(t)}_{\hat{\sigma}}$ is the component of $\Ut_i^{(t)}$ that corresponds to sequence $\hat{\sigma}$;
    \item[(iii)] $\vS_{\hat{\sigma}}^{(t-1)} \defeq \tilde{\Ut}_{\hat{\sigma}}^{(t-1)} + \sum_{\tau=1}^{t-1} \Tilde{\Ut}_{\hat{\sigma}}^{(\tau)}$; and
    \item[(iv)] $\eta > 0$ is the learning rate---common among all $\regdep_{\hat{\sigma}}$.
\end{itemize}

Finally, having determined $\tilde{\vq}^{(t)}_{\hat{\sigma}} = (\lambda^{(t)}_{\hat{\sigma}}, \vy^{(t)}_{\hat{\sigma}})$, we compute $\vq_{\hat{\sigma}}^{(t)} \defeq \frac{\vy^{(t)}_{\hat{\sigma}}}{\lambda^{(t)}_{\hat{\sigma}}} \in \cQ_j$. In turn, this gives the next strategy of $\regdep_{\hat{\sigma}}$ as $\vX^{(t)}_{\hat{\sigma}} \defeq \mat{M}_{\hat{\sigma} \to \vq^{(t)}_{\hat{\sigma} }}$ (recall \Cref{def:matM}). It is evident that the regret minimization problem faced by each $\regdep_{\hat{\sigma}}$ is equivalent to minimizing regret over $\cQ_j$, since only the components of $\vX^{(t)}_{\hat{\sigma}}$ that correspond to $\vq_{\hat{\sigma}}^{(t)}$ cumulate regret (the rest are constant), leading to the regret bound below. We note that all the subsequent analysis operates under the tacit premise that each local regret minimizer is updated via~\algoshort, without explicitly mentioned in the statements in order to lighten the exposition.

\begin{proposition}
    \label{prop:orig-rvu}
    For any learning rate $\eta \leq \frac{1}{256 \|\cQ_i\|_1}$ and $T \geq 2$, $\max\{0, \reg^T_{\hat{\sigma}} \}$ can be upper bounded by
    \begin{align*}
        2 \frac{|\Sigma_i| \log T}{\eta} &+ 16 \eta \|\cQ_i\|^2 \sum_{t=1}^{T-1} \|\Ut_i^{(t+1)} - \Ut_i^{(t)} \|_\infty^2 - \frac{1}{32\eta} \sum_{t=1}^{T-1} \left\| \begin{pmatrix}
            \lambda_{\hat{\sigma}}^{(t+1)} \\
            \vy_{\hat{\sigma}}^{(t+1}
        \end{pmatrix} -
        \begin{pmatrix}
            \lambda_{\hat{\sigma}}^{(t)} \\
            \vy_{\hat{\sigma}}^{(t}
        \end{pmatrix}
        \right\|^2_{t}.
    \end{align*}
\end{proposition}

\begin{proof}
    This regret bound is an immediate implication of~\citep[Proposition 2 and Corollary 1]{Farina22:Near}. More precisely, we note that the regret bound by~\citet{Farina22:Near} applies if $\|\Ut_i^{(t)}\|_\infty \leq \frac{1}{\|\cQ_i\|_1}$, for any $t \in \N$. That assumption can be met by rescaling the learning rate by a factor of $\frac{1}{\|\cQ_i\|_1}$ since in our setting it holds that $\|\Ut_i^{(t)}\|_\infty \leq 1$; the latter follows from the definition of $\Ut_i^{(t)} \defeq \ut_i^{(t)} \otimes \vx_i^{(t)}$ (\Cref{line:outer}), and the fact that $\|\ut_i^{(t)}\|_\infty \leq 1$ (by assumption) and $\|\vx_i^{(t)}\|_\infty \leq 1$ (since $\cQ_i \subseteq [0,1]^{|\Sigma_i|}$).
\end{proof}
In \Cref{prop:orig-rvu} we used the shorthand notation
\begin{equation*}
    \left\| \begin{pmatrix}
            \lambda_{\hat{\sigma}}^{(t+1)} \\
            \vy_{\hat{\sigma}}^{(t+1}
        \end{pmatrix} -
        \begin{pmatrix}
            \lambda_{\hat{\sigma}}^{(t)} \\
            \vy_{\hat{\sigma}}^{(t}
        \end{pmatrix}
        \right\|^2_{t} \defeq \left\| \begin{pmatrix}
            \lambda_{\hat{\sigma}}^{(t+1)} \\
            \vy_{\hat{\sigma}}^{(t+1}
        \end{pmatrix} -
        \begin{pmatrix}
            \lambda_{\hat{\sigma}}^{(t)} \\
            \vy_{\hat{\sigma}}^{(t}
        \end{pmatrix}
        \right\|^2_{(\lambda^{(t)}_{\hat{\sigma}}, \vy^{(t)}_{\hat{\sigma}})},
\end{equation*}
where for a vector $\tilde{\vec{w}} \in \R^{d+1}$ and $\tilde{\vx} \in \R^{d+1}_{> 0}$, we used the notation
\begin{equation*}
    \|\tilde{\vec{w}}\|_{\tilde{\vx}} \defeq \sqrt{\sum_{\edg \in \range{d+1}} \left( \frac{\tilde{\vec{w}}[\edg]}{\tilde{\vx}[\edg]} \right)^2 }
\end{equation*}
for the local norm induced by $\tilde{\vx}$. Further, we will also use the notation
\begin{equation*}
    \|\tilde{\vec{w}}\|_{\tilde{\vx},\infty} \defeq \max_{\edg \in \range{d+1}} \left| \frac{\tilde{\vec{w}}[\edg]}{\tilde{\vx}[\edg]} \right|.
\end{equation*}

\begin{lemma}
    \label{lemma:q-mul-stab}
    For any sequence $\hat{\sigma} = (j,a) \in \Sigma_i^*$, learning rate $\eta \leq \frac{1}{50 \|\cQ_i\|_1}$ and time $t \in \range{T-1}$,
    \begin{align*}
        \max_{\sigma \in \Sigma_j} \left| 1 - \frac{\vq_{\hat{\sigma}}^{(t+1)}[\sigma]}{\vq_{\hat{\sigma}}^{(t)}[\sigma]} \right| &\leq 4 \left\| \begin{pmatrix}
            \lambda_{\hat{\sigma}}^{(t+1)} \\
            \vy_{\hat{\sigma}}^{(t+1)}
        \end{pmatrix} - 
        \begin{pmatrix}
            \lambda_{\hat{\sigma}}^{(t)} \\
            \vy_{\hat{\sigma}}^{(t)}
        \end{pmatrix}
        \right\|_{t, \infty} \leq 100 \eta \|\cQ_i\|_1.
    \end{align*}
\end{lemma}

\begin{proof}
    We will need the following stability bound, extracted from~\citep[Proposition 3]{Farina22:Near}.
\begin{lemma}[\nakedcite{Farina22:Near}]
    \label{lemma:Farina-stab}
    For any sequence $\hat{\sigma} = (j,a) \in \Sigma_i^*$, time $t \in \range{T-1}$ and learning rate $\eta \leq \frac{1}{50 \|\cQ_i\|_1}$,
    \begin{equation*}
        \left\| \begin{pmatrix}
            \lambda_{\hat{\sigma}}^{(t+1)} \\
            \vy_{\hat{\sigma}}^{(t+1)}
        \end{pmatrix} - 
        \begin{pmatrix}
            \lambda_{\hat{\sigma}}^{(t)} \\
            \vy_{\hat{\sigma}}^{(t)}
        \end{pmatrix}
        \right\|_{t, \infty} \leq 22 \eta \|\cQ_i\|_1.
    \end{equation*}
\end{lemma}
Now let us fix a time $t \in \range{T-1}$. For convenience, we introduce the notation
    \begin{equation}
        \mul^{(t)} \defeq \left\| \begin{pmatrix}
            \lambda_{\hat{\sigma}}^{(t+1)} \\
            \vy_{\hat{\sigma}}^{(t+1)}
        \end{pmatrix} - 
        \begin{pmatrix}
            \lambda_{\hat{\sigma}}^{(t)} \\
            \vy_{\hat{\sigma}}^{(t)}
        \end{pmatrix}
        \right\|_{t, \infty}.
    \end{equation}
    For our choice of the learning rate $\eta \leq \frac{1}{50\|\cQ_i\|_1}$, \Cref{lemma:Farina-stab} implies that $\mul^{(t)} \leq \frac{1}{2}$.
    By definition, we have
    \begin{align*}
        \vq_{\hat{\sigma}}^{(t+1)} \defeq \frac{\vy_{\hat{\sigma}}^{(t+1)}}{\lambda_{\hat{\sigma}}^{(t+1)}} &\leq \frac{(1 + \mul^{(t)}) \vy_{\hat{\sigma}}^{(t)}}{(1 - \mul^{(t)}) \lambda_{\hat{\sigma}}^{(t)}} = \left( 1 + \frac{2\mul^{(t)}}{1 - \mul^{(t)}} \right) \vq_{\hat{\sigma}}^{(t)} \leq (1 + 4 \mul^{(t)}) \vq_{\hat{\sigma}}^{(t)},
    \end{align*}
    where the last bound follows since $\mul^{(t)} \leq \frac{1}{2}$. That is,
    \begin{equation}
        \label{eq:upper-X-dev}
        \frac{\vq_{\hat{\sigma}}^{(t+1)}}{\vq_{\hat
        \sigma}^{(t)}} \leq (1 + 4 \mul^{(t)}).
    \end{equation}
    Similarly,
    \begin{align*}
        \vq_{\hat{\sigma}}^{(t+1)} = \frac{\vy_{\hat{\sigma}}^{(t+1)}}{\lambda_{\hat{\sigma}}^{(t+1)}} \geq \frac{1 - \mul^{(t)}}{1 + \mul^{(t)}} \frac{\vy_{\hat{\sigma}}^{(t)}}{\lambda_{\hat{\sigma}}^{(t)}} = \left( 1 - \frac{2\mul^{(t)}}{1 + \mul^{(t)}} \right) \vq_{\hat{\sigma}}^{(t)} \geq (1 - 2 \mul^{(t)}) \vq_{\hat{\sigma}}^{(t)}.
    \end{align*}
    Thus,
    \begin{equation}   
        \label{eq:lower-X-dev}
        \frac{\vq_{\hat{\sigma}}^{(t+1)}}{\vq_{\hat{\sigma}}^{(t)}} \geq 1 - 2 \mul^{(t)}.
    \end{equation}
    As a result, the claim follows from \eqref{eq:upper-X-dev} and \eqref{eq:lower-X-dev}.
\end{proof}
We are now ready to establish \Cref{prop:rvu-sigma}, restated below.

\rvusigma*

\begin{proof}
    The claim follows directly from \Cref{prop:orig-rvu} and \Cref{lemma:q-mul-stab}.
\end{proof}

Under the premise that $\regdep_{\triangle}$ is also updated via~\algoshort, similar reasoning yields the proof of \Cref{prop:rvu-tri}.

\rvutri*

\begin{proof}
    The argument is analogous to the proof of \Cref{prop:rvu-sigma}, leveraging the fact that $\| \uttri^{(t)}\|_\infty = |\langle \vX^{(t)}_{\hat{\sigma}}, \Ut^{(t)}_i \rangle| \leq \|\vX^{(t)}_{\hat{\sigma}}\|_1 \|\Ut_i^{(t)}\|_\infty \leq 2 |\Sigma_i|$, for any $\hat{\sigma} \in \Sigma_i^*$, by Cauchy-Schwarz inequality.
\end{proof}

\begin{lemma}
    \label{lemma:l-mul-sta}
    For any $t \in \range{T-1}$ and $\etatri \leq \frac{1}{512|\Sigma_i|}$,
    \begin{equation*}
        \max_{\hat{\sigma} \in \Sigma_i^*} \left| 1 - \frac{\vlam_i^{(t+1)}[\hat{\sigma}]}{\vlam_i^{(t)}[\hat{\sigma}]} \right| \leq 200 \etatri |\Sigma_i|.
    \end{equation*}
\end{lemma}

\begin{proof}
    The argument is analogous to \Cref{lemma:q-mul-stab}.
\end{proof}

Next, we combine all those local regret minimizers, namely $\regdep_{\triangle}, (\regdep_{\hat{\sigma}})_{\hat{\sigma} \in \Sigma_i^*}$, into a global regret minimizer $\regdep_{\Psi_i}$ for the set $\Psi_i$ via the regret circuit for the convex hull. Finally, we denote by $\regdep$ the $\Psi_i$-regret minimizer derived from \Cref{algo:Gordon}, based on $\regdep_{\Psi_i}$.

\hull*

\begin{proof}
    Using the guarantee of the regret circuit for the convex hull~\citep{Farina19:Regret},
    we have
    \begin{equation*}
        \reg^T \leq  \reg_{\triangle}^T + \max_{\hat{\sigma} \in \Sigma_i^*} \reg^T_{\hat{\sigma}},
    \end{equation*}
    where $\reg^T$ is the external regret cumulated by $\regdep_{\Psi_i}$ up to time $T$. But, by \Cref{theorem:gordon}, this is equal to the $\Psi_i$-regret of $\regdep$, constructed according to \Cref{algo:Gordon}. As a result,
    \begin{equation*}
        \reg_{\Psi_i}^T \leq  \reg_{\triangle}^T + \max_{\hat{\sigma} \in \Sigma_i^*} \reg^T_{\hat{\sigma}},
    \end{equation*}
    In turn, this implies that
    \begin{align*}
        \max\{0, \reg^T_{\Psi_i}\} &\leq \max \left\{0, \reg_{\triangle}^T + \max_{\hat{\sigma} \in \Sigma_i^*} \reg^T_{\hat{\sigma}}  \right\} \\
        &\leq \max\{0, \reg_{\triangle}^T\} + \max_{\hat{\sigma} \in \Sigma_i^*} \max\{0, \reg^T_{\hat{\sigma}}\} \\
        &\leq \max\{0, \reg_{\triangle}^T\} + \sum_{\hat{\sigma} \in \Sigma_i^*} \max\{0, \reg_{\hat{\sigma}}^T \},
    \end{align*}
    where the last inequality follows from the fact that $\max\{0, \reg^T_{\hat{\sigma}}\} \geq 0$, for any $\hat{\sigma} \in \Sigma_i^*$.
\end{proof}

\subsection{Characterizing the Fixed Points}

We recall that $(\vx_i^{(t)})_{1 \leq t \leq T}$ denotes the sequence of fixed points produced by \Cref{algo:Gordon}---that is, the strategies produced by $\regdep$. The next key result relates the deviation of the fixed points---in $\ell_1$ norm---in terms of the \emph{multiplicative deviation} of the transformations, assuming a particular rational function characterization of the fixed points.

\gamma*

\begin{proof}
    Let us define
    \begin{equation}
        \label{eq:defmul}
        \mul^{(t)} \defeq \max_{\edg \in \range{D}} \left| 1 - \frac{\vX_i^{(t+1)}[\edg]}{\vX_i^{(t)}[\edg]} \right|.
    \end{equation}
    By assumption, it holds that $\mul^{(t)} \leq \frac{100}{256\deg_i} \leq \frac{1}{2\deg_i}$. Further, suppose that
    \begin{equation}
        \label{eq:p-poly}
        p_{\sigma,k} : \vX_i \mapsto \sum_{\tree \in \treeset_{\sigma, k}} C_{\tree} \prod_{\edg \in \tree} \vX_i[\edg],
    \end{equation}
    and
    \begin{equation}
        \label{eq:q-poly}
        q_{\sigma, k}: \vX_i \mapsto \sum_{\tree \in \treeset'_{\sigma, k}} C_{\tree} \prod_{\edg \in \tree} \vX_i[\edg],
    \end{equation}
    for all $(\sigma, k) \in \Sigma_i \times \range{m}$, where $C_{\tree} > 0$ for any $\tree \in \treeset_{\sigma, k}$ and $C_{\tree} > 0$ for any $\tree \in \treeset'_{\sigma, k}$. Here, $\tree$ can be a multiset or an empty set (the validity of~\eqref{eq:p-poly} and \eqref{eq:q-poly} follows by assumption). Then, for $(\sigma,k) \in \Sigma_i \times \range{m}$,
    \begin{align}
        p_{\sigma, k}(\vX_i^{(t+1)}) &= \sum_{\tree \in \treeset_{\sigma, k}} C_{\tree} \prod_{\edg \in \tree} \vX_i^{(t+1)}[\edg] \notag \\
        &\leq \sum_{\tree \in \treeset_{\sigma, k}} C_{\tree} \prod_{\edg \in \tree} (1 + \mul^{(t)}) \vX_i^{(t)}[\edg] \label{align:mulX} \\
        &\leq (1 + \mul^{(t)})^{\deg_i} \sum_{\tree \in \treeset_{\sigma, k}} C_{\tree} \prod_{\edg \in \tree} \vX_i^{(t)}[\edg] \label{align:deg} \\
        &= (1 + \mul^{(t)})^{\deg_i} p_{\sigma, k}(\vX_i^{(t)}) \notag \\
        &\leq (1 + 1.5 \mul^{(t)} \deg_i) p_{\sigma, k}(\vX_i^{(t)}), \label{align:calc}
    \end{align}
    where \eqref{align:mulX} follows since $\vX_i^{(t+1)}[\edg] \leq (1 + \mul^{(t)}) \vX_i^{(t)}[\edg]$, for any $\edg \in \range{D}$, by definition of $\mul^{(t)}$ in \eqref{eq:defmul}; \eqref{align:deg} uses the fact that $|\tree| \leq \deg$ for any $\tree \in \treeset_{\sigma, k}$; and \eqref{align:calc} follows since $(1 + \mul^{(t)})^{\deg_i} \leq \exp \{ \mul^{(t)} \deg_i \} \leq 1 + 1.3 \mul^{(t)} \deg_i$ for $\mul^{(t)} \leq \frac{1}{2 \deg_i}$. Similarly, for $(\sigma, k) \in \Sigma_i \times \range{m}$, we get
    \begin{align}
        p_{\sigma, k}(\vX_i^{(t+1)}) &= \sum_{\tree \in \treeset_{\sigma, k}} C_{\tree} \prod_{\edg \in \tree} \vX_i^{(t+1)}[\edg] \notag \\
        &\geq \sum_{\tree \in \treeset_{\sigma, k}} C_{\tree} \prod_{\edg \in \tree} (1 - \mul^{(t)}) \vX_i^{(t)}[\edg] \notag \\
        &\geq (1 - \mul^{(t)})^{\deg_i} p_{\sigma, k}(\vX_i^{(t)}) \notag \\
        &\geq (1 - \mul^{(t)} \deg_i) p_{\sigma, k}(\vX_i^{(t)}),\label{align:finalp}
    \end{align}
    where the last bound follows from Bernoulli's inequality. Analogous reasoning yields that for any $(\sigma, k) \in \Sigma_i \times \range{m}$,
    \begin{equation}
        \label{eq:q1}
        q_{\sigma, k}(\vX_i^{(t+1)}) \leq (1 + 1.3 \mul^{(t)} \deg_i) q_{\sigma, k}(\vX_i^{(t)}),
    \end{equation}
    and 
    \begin{equation}
        \label{eq:q2}
        q_{\sigma, k}(\vX_i^{(t+1)}) \geq (1 - \mul^{(t)} \deg_i) q_{\sigma, k}(\vX_i^{(t)}).
    \end{equation}
    As a result, for $\sigma \in \Sigma_i$,
    \begin{align}
        \vx_i^{(t+1)}[\sigma] - \vx_i^{(t)}[\sigma] &= \sum_{k=1}^m \frac{p_{\sigma, k}(\vX_i^{(t+1)})}{q_{\sigma, k}(\vX_i^{(t+1)})} - \sum_{k=1}^m \frac{p_{\sigma, k}(\vX_i^{(t)})}{q_{\sigma, k}(\vX_i^{(t)})}  \notag \\
        &\leq \sum_{k=1}^m \left( \frac{1 + 1.3 \mul^{(t)} \deg_i}{1 - \mul^{(t)} \deg_i} \right) \frac{p_{\sigma, k}(\vX^{(t)})}{q_{\sigma, k}(\vX_i^{(t)})} - \sum_{k=1}^m \frac{p_{\sigma, k}(\vX_i^{(t)})}{q_{\sigma, k}(\vX_i^{(t)})} \label{align:updown} \\
        &\leq \left( 1 + \frac{2.3 \mul^{(t)}\deg_i}{1 - \mul^{(t)} \deg} \right) \sum_{k=1}^m \frac{p_{\sigma, k}(\vX_i^{(t)})}{q_{\sigma, k}(\vX_i^{(t)})} - \sum_{k=1}^m \frac{p_{\sigma, k}(\vX_i^{(t)})}{q_{\sigma, k}(\vX_i^{(t)})} \notag \\
        &= \frac{2.3 \mul^{(t)}\deg_i}{1 - \mul^{(t)} \deg_i} \vx_i^{(t)}[\sigma] \leq 4 \mul^{(t)} \deg_i \vx_i^{(t)}[\sigma]. \label{align:onedif}
    \end{align}
    where \eqref{align:updown} uses \eqref{align:calc} and \eqref{eq:q2}, and \eqref{align:onedif} follows from the fact that $\mul^{(t)} \leq \frac{100}{256 \deg_i}$. Similarly, by \eqref{align:finalp} and \eqref{eq:q1},
    \begin{align*}
        \vx_i^{(t)}[\sigma] - \vx_i^{(t+1)}[\sigma] &= \sum_{k=1}^m \frac{p_{\sigma, k}(\vX_i^{(t)})}{q_{\sigma, k}(\vX_i^{(t)})} - \sum_{k=1}^m \frac{p_{\sigma, k}(\vX_i^{(t+1)})}{q_{\sigma, k}(\vX_i^{(t+1)})} \\ 
        &\leq 4 \mul^{(t)} \deg_i \vx_i^{(t)}[\sigma].
    \end{align*}
    As a result, we conclude that
    \begin{align*}
        \| \vx_i^{(t+1)} - \vx_i^{(t)} \|_1 &\leq 4 \mul^{(t)} \deg_i \|\cQ_i\|_1.
    \end{align*}
\end{proof}

\Cref{theorem:rational} makes the assumption that each polynomial in~\eqref{eq:rational} has positive coefficients. While this might seem rather restrictive, we next show that there is a procedure that eliminates the negative monomials, as long as the involved variables are deriving from the sequence-form polytope. As a warm-up, we first establish this property for variables deriving from the simplex. 

We note that the processes described in the proofs below are not meant to be algorithmic meaningful, but instead highlight the generality of \Cref{theorem:rational}. Indeed, the way one computes the fixed point should not be necessarily related to the rational function formula postulated in~\eqref{eq:rational}; for example, \emph{computing} the stationary distribution of a Markov chain using the Markov chain tree theorem would make little sense, as it would require exponential time.

\begin{proposition}
    \label{prop:poscoeffs}
    Let $p : \vX \mapsto \R$ be a non-constant multivariate polynomial of degree $\deg \in \N$. If $\vX = (\vx_1, \dots, \vx_m)$ such that $\vx_k \in \Delta^{d_k}$, for all $k \in \range{m}$, $p$ can be expressed as a combination of monomials with positive coefficients of degree at most $\deg$ and a constant term.\footnote{We thank Kaito Fujii for pointing out an inconsistency in \Cref{prop:poscoeffs,prop:poscoeffs-tree} present in an earlier version.}
\end{proposition}

\begin{proof}
    Let
    \begin{equation*}
        p(\vX) = \sum_{\tree \in \treeset} C_\tree \prod_{\edg \in \tree} \vX[\edg] + C,
    \end{equation*}
    where $\treeset$ is a finite and nonempty set, and $\tree \neq \emptyset$ and $C_{\tree} \neq 0$ for all $\tree \in \treeset$. To establish the claim, we consider the following iterative algorithm. 

    First, if it happens that $C_{\tree} > 0$, for all $\tree \in \treeset$, the algorithm terminates. Otherwise, we take any monomial of the form $C_{\tree} \prod_{e \in \tree} \vX[\edg]$ with $C_{\tree} < 0$. Since $\tree \neq \emptyset$, we might take $\edg \in \tree$. Further, we let $\vX[\edg] = \vx_k[\ind]$, for some $k \in \range{m}, \ind \in \range{d_k}$, where $\vx_k \in \Delta^{d_k}$. As such, we have that $\vx_k[\ind] = 1 - \sum_{\ind' \neq \ind} \vx_k[\ind']$. Thus,
    \begin{align*}
        C_{\tree} \prod_{\edg' \in \tree} \vX[\edg'] = C_{\tree} \prod_{\edg' \in \tree \setminus \{\edg\}} \vX[\edg'] 
        + \sum_{\ind' \neq \ind} (- C_{\tree}) \vx_k[\ind'] \prod_{\edg' \in \tree \setminus \{\edg\}} \vX[\edg'].
    \end{align*}
    Here, by convention the product over an empty set is assumed to be $1$. This step clearly cannot increase the degree of the polynomial. Now to analyze this iterative process, we consider as the potential function the sum of the degrees of all the negative monomials---monomials for which $C_{\tree} < 0$. It should be evident that every step of the previous algorithm will decrease the potential function by one. Further, the previous step can always be applied as long as the potential function is not zero. As a result, given that $\treeset$ is finite, we conclude that after a finite number of iterations the potential function will be zero. Then, we will have that
    \begin{equation*}
            p(\vX) = \sum_{\tree \in \treeset'} C_{\tree} \prod_{e \in \tree} \vX[\edg] + C',
    \end{equation*}
    where $\tree \neq \emptyset$, $C_{\tree} > 0$, and $C' \in \R$. This concludes the proof.
\end{proof}

\begin{proposition}
    \label{prop:poscoeffs-tree}
    Let $p : \vX \mapsto \R$ be a non-constant multivariate polynomial of degree $\deg \in \N$. If $\vX = (\vq_1, \dots, \vq_m)$ such that $\vq_k \in \cQ^{d_k}$, for all $k \in \range{m}$, $p$ can be expressed as a combination of monomials with positive coefficients of degree at most $\deg$ and a constant term.
\end{proposition}

\begin{proof}
    As in \Cref{prop:poscoeffs}, let
    \begin{equation*}
        p(\vX) = \sum_{\tree \in \treeset} C_\tree \prod_{\edg \in \tree} \vX[\edg] + C,
    \end{equation*}
    where $\treeset$ is a finite and nonempty set, and $\tree \neq \emptyset$ and $C_{\tree} \neq 0$ for all $\tree \in \treeset$. We consider the following algorithm. 

    First, if $C_{\tree} > 0$, for all $\tree \in \treeset$, the algorithm may terminate. In the contrary case, we consider any monomial $C_{\tree} \prod_{e \in \tree} \vX[\edg]$ for which $C_{\tree} < 0$. Further, take any $\edg \in \tree$, which is possible since $\tree \neq \emptyset$. Now let us assume that $\vX[\edg] = \vq_k[\sigma]$, for some $k \in \range{m}, \sigma = (j,a)$. By the sequence-form polytope constraints, we have
    \begin{equation*}
        \vq_k[\sigma] = \vq_k[\sigma_j] - \sum_{a' \in \cA_j \setminus \{a\}} \vq_k[(j,a')].
    \end{equation*}
    Thus,
    \begin{align*}
        C_{\tree} \prod_{\edg' \in \tree} \vX[\edg'] = C_{\tree} \vq_k[\sigma_j] \prod_{\edg' \in \tree \setminus \{\edg\}} \vX[\edg'] + \sum_{a' \neq a} (- C_{\tree}) \vq_k[(j, a')] \prod_{\edg' \in \tree \setminus \{\edg\}} \vX[\edg'].
    \end{align*}
    This step clearly does not increase the degree of the polynomial. To construct a potential function, we will say that the \emph{depth} of a monomial $\prod_{\edg \in \tree} \vX[\edg]$, for $\tree \neq \emptyset$, is the sum of the depths of each $\vX[\edg]$; more precisely, the depth of $\vq_k[\sigma]$ is $0$ if $\sigma = \emptyseq$, or $1$ plus the depth of $\vq_k[\sigma_j]$ otherwise. Now we claim that the sum of the depths of the negative monomials is a proper potential function. Indeed, by construction every step reduces the potential by $1$, while the previous step can always be applied when the potential function is not zero. As a result, given that $\treeset$ is finite, we conclude that after a finite number of iterations the potential function will be zero, which in turn implies that
    \begin{equation*}
        p(\vX) = \sum_{\tree \in \treeset'} C_{\tree} \prod_{\edg \in \tree} \vX[\edg] + C',
    \end{equation*}
    where $\tree \neq \emptyset$, $C_{\tree} > 0$, and $C' \in \R$. This concludes the proof.
\end{proof}

The main caveat of \Cref{prop:poscoeffs,prop:poscoeffs-tree} is that the constant term in the resulting polynomials could be negative, an issue we do not address in this work.

Now we show that the fixed points associated with EFCCE and EFCE can be analyzed through the lens of \Cref{theorem:rational}, establishing \Cref{prop:fp-efcce,prop:fp-efce}. 

\efccerational*

\begin{proof}
    Consider any coarse trigger deviation function $\phi^{(t)}_i = \sum_{j \in \cJ_i} \vlam^{(t)}_i[j] \phi_{j \to \vq^{(t)}_j}$, where $\vq^{(t)}_j \in \cQ_j$ (\Cref{claim:equiv-efcce}). Given that we are updating $\regdep_{\triangle}$ using \algoshort, it follows that $\vlam_i^{(t)}[j] > 0$ for any $j \in \cJ_i$. As a result, by~\citep[Theorem 5.1]{Anagnostides22:Faster}, the (unique) fixed point $\vx^{(t)}_i \in \cQ_i$ can be computed in a top-down fashion as follows.
    \begin{equation}
        \label{eq:realind}
        \vx^{(t)}_i[\sigma] = \frac{\sum_{j' \preceq j} \vlam^{(t)}_i[j'] \vq^{(t)}_{j'}[\sigma] \vx^{(t)}_i[\sigma_{j'}]}{\sum_{j' \preceq j} \vlam^{(t)}_i[j']},
    \end{equation}
    for any sequence $\sigma = (j,a) \in \Sigma_i^*$. We will prove the claim by induction. For the basis of the induction, we note that the empty sequence is trivially given by a $0$-degree rational function with positive coefficients; namely, $\vx_i[\emptyseq] = \frac{1}{1}$.
    
    Now for the inductive step, let us take any sequence $\sigma = (j,a) \in \Sigma_i^*$. We suppose that for any sequence $\sigma_{j'}$, for $j' \preceq j$, it holds that
    \begin{equation*}
        \vx^{(t)}_i[\sigma_{j'}] = \sum_{k=1}^{m_{\sigma_{j'}}} \frac{p_{\sigma_{j'}, k}(\vX_i^{(t)})}{q_{\sigma_{j'}, k}(\vX_i^{(t)})},
    \end{equation*}
    where $\{p_{\sigma_{j'}, k}\}, \{q_{\sigma_{j'}, k}\}$ are multivariate polynomials with positive coefficients and maximum degree at most $h \in \N \cup \{0\}$. Then, the term
    \begin{equation*}
        \frac{ \vlam_i^{(t)}[j'] \vq^{(t)}_{j'}[\sigma]}{  \sum_{j' \preceq j} \vlam_i^{(t)}[j']} \cdot \frac{p_{\sigma_{j'}, k}(\vX_i^{(t)})}{q_{\sigma_{j'}, k}(\vX_i^{(t)})}
    \end{equation*}
    is a rational function in $\vX_i^{(t)}$ with positive coefficients and maximum degree at most $h + 2$. Hence, the term below is a sum of rational functions with positive coefficients and maximum degree at most $h + 2$:
    \begin{equation*}
        \frac{\sum_{j' \preceq j} \vlam^{(t)}_i[j'] \vq^{(t)}_{j'}[\sigma] \vx^{(t)}_i[\sigma_{j'}]}{\sum_{j' \preceq j} \vlam^{(t)}_i[j']}
    \end{equation*}
    As a result, by~\eqref{eq:realind} we conclude that 
    \begin{equation*}
     \vx^{(t)}_i[\sigma] = \sum_{k=1}^{m_{\sigma}} \frac{p_{\sigma, k}(\vX_i^{(t)})}{q_{\sigma, k}(\vX_i^{(t)})}, 
    \end{equation*}
    where $\{p_{\sigma, k}\}, \{q_{\sigma ,k}\}$ have positive coefficients and maximum degree $h + 2$. This establishes the inductive step, concluding the proof.
\end{proof}

Next, for the proof of \Cref{prop:fp-efce}, we will need the following key refinement of the Markov chain tree theorem~\citep[Corollary A.8]{Anagnostides22:Faster}.

\begin{theorem}[\nakedcite{Anagnostides22:Faster}]
    \label{theorem:rankone}
    Let $\mat{M}$ be the transition matrix of a $d$-state Markov chain such that $\mat{M} = \Vec{v} \Vec{1}_{d}^\top + \mat{C}$, where $\mat{C} \in \R_{> 0}^{d \times d}$ and $\Vec{v} \in \R_{> 0}^d$ has entries summing to $\lambda > 0$. Further, let $\Vec{v} = \Vec{r}/l$, for some $l > 0$. If $\vx \in \Delta^d$ is the (unique) stationry distribution of $\mat{M}$, then for each $\ind \in \range{d}$ there exist a nonempty and finite set $F_{\ind}$, and $F = \cup_{\ind=1}^d F_{\ind}$, and parameters $b_k \in \{0,1\}$, $0 \leq p_k \leq d - 2$, $|S_k| = d - p_k - b_k - 1$, for each $k \in F_{\ind}$, such that the $\ind$-th coordinate of $\vec{w} \defeq l \Vec{x}$ can be expressed as 
    \begin{equation*}
        \vec{w}[\ind] = \frac{\sum_{k \in F_{\ind}} \lambda^{p_k + 1} (\vec{r}[q_k])^{b_k} l^{1 - b_k} \prod_{(a,b) \in S_k} \mat{C}[a,b]}{\sum_{k \in F} \lambda^{p_k + b_k} C_k \prod_{(a,b) \in S_k} \mat{C}[a,b]},
    \end{equation*}
    for each $\ind \in \range{d}$, where $C_k = C_k(d) > 0$.
\end{theorem}

Let us also introduce the following terminology, borrowed from the work of~\citet{Farina21:Simple}.

\begin{definition}[\nakedcite{Farina21:Simple}]
   \label{def:trunk}
   Let $J \subseteq \cJ_i$ be a subset of $i$'s information sets. We say that $J$ is a trunk of $\cJ_i$ if for all $j \in J$, all predecessors of $j$ are also in $J$.
\end{definition}

\begin{definition}[\nakedcite{Farina21:Simple}]
    \label{def:partial}
    Let $\phi_i \in \Psi_i$ and $J$ be a trunk of $\cJ_i$. We say that a vector $\vx_i \in \R_{\geq 0}^{|\Sigma_i|}$ is a $J$-partial fixed point if it satisfies all the sequence-form constraints at all information sets $j \in \cJ_i$, and
    \begin{equation*}
        \phi_i(\vx_i)[\emptyseq] = \vx_i[\emptyseq] = 1,
    \end{equation*}
    \begin{equation*}
        \phi_i(\vx_i)[(j,a)] = \vx_i[(j,a)], \quad \forall j \in \cJ_i, a \in \cA_j.
    \end{equation*}
\end{definition}

\rationalefce*

\begin{proof}
    For the base of the induction, the claim trivially holds for $\vx^{(t)}_i[\emptyseq] = \frac{1}{1}$. For the inductive step, let us first define a vector $\vr^{(t)} \in \R^{|\cA_{j^*}|}_{\geq 0}$, so that $\vr^{(t)}[a]$ is equal to
    \begin{equation*}
        \sum_{j' \preceq \sigma_{j^*}} \sum_{a' \in \cA_{j'}} \vlam^{(t)}_i[(j',a')] \vq^{(t)}_{(j',a')}[(j^*,a)]  \vx^{(t)}_i[(j', a')].
    \end{equation*}
    Further, we let $\mat{W}^{(t)} \in \mathbb{S}^{|\cA_{j^*}|}$ be a stochastic matrix, so that for any $a_r, a_c \in \cA_{j^*}$, $\mat{W}^{(t)}[a_r, a_c]$ is equal to
    \begin{align}
        \frac{1}{\vx^{(t)}_i[\sigma_{j^*}]} \vr^{(t)}[a_r] + \vlam_i^{(t)}[(j^*, a_c)] \vq_{(j^*, a_c)}^{(t)}[(j^*, a_r)] + \left( 1 -  \sum_{\hat{\sigma} \preceq (j^*, a_c)} \vlam_i^{(t)}[\hat{\sigma}] \right) \mathbbm{1}\{a_r = a_c\},\notag
    \end{align}
    By \citep[Proposition 4.14]{Farina21:Simple}, if $\vb^{(t)} \in \Delta(\cA_{j^*})$ is the (unique) stationary distribution of $\mat{W}^{(t)}$, extending by $\vx_i^{(t)}[\sigma_{j^*}] \vb^{(t)}$ at information set $j^*$ yields a $(J \cup \{j^* \})$-partial fixed point (\Cref{def:partial}). To bound the increase in the degree of the rational function, we will use \Cref{theorem:rankone}. In particular, we define a matrix $\mat{C}^{(t)} \in \R^{|\cA_{j^*}| \times |\cA_{j^*}|}$, so that for any $a_r, a_c \in \cA_{j^*}$,
    \begin{align}
        \mat{C}^{(t)}[a_r, a_c] \defeq \vlam_i^{(t)}[(j^*, a_c)] \vq_{(j^*, a_c)}^{(t)}[(j^*, a_r)] + \left( 1 -  \sum_{\hat{\sigma} \preceq (j^*, a_c)} \vlam_i^{(t)}[\hat{\sigma}] \right) \mathbbm{1}\{a_r = a_c\}. \label{eq:matC}
    \end{align}
    For a fixed $a_c \in \cA_{j^*}$, we have
    \begin{equation}
        \label{eq:sumC}
        \sum_{a_r \in \cA_{j^*}} \!\!\!\!\! \mat{C}^{(t)}[a_r, a_c] = \vlam^{(t)}_i[(j^*, a_c)] + \!\!\!\!\!\! \sum_{\hat{\sigma} \not\preceq (j^*, a_c)} \!\!\!\!\! \vlam^{(t)}_i[\hat{\sigma}],
    \end{equation}
    where we used the fact that for any $a_c \in \cA_{j^*}$, $\sum_{a_r \in \cA_{j^*}} \vq_{(j^*, a_c)}[(j^*, a_r)] = 1$ since $\vq_{(j^*, a_c)} \in \cQ_{j^*}$. Thus, from \eqref{eq:sumC} we obtain that
    \begin{equation}
        \label{eq:lam}
        1 - \sum_{a_r \in \cA_{j^*}} \mat{C}^{(t)}[a_r, a_c] = \sum_{\hat{\sigma} \prec (j^*, a_c)} \vlam^{(t)}_i[\hat{\sigma}].
    \end{equation}
    Now for the inductive step, suppose that for any information set $j' \preceq \sigma_{j^*}$ and $a' \in \cA_{j'}$, the partial fixed point $\vx^{(t)}_i[\sigma']$, with $\sigma' = (j', a')$, can be expressed as
    \begin{equation}
        \label{eq:ind-hyp}
        \vx^{(t)}_i[\sigma'] = \sum_{k=1}^{m_{\sigma'}} \frac{p_{\sigma', k}(\vX_i^{(t)})}{q_{\sigma', k} (\vX_i^{(t)})},
    \end{equation}
    where $\{p_{\sigma', k}\}, \{q_{\sigma', k}\}$ are multivariate polynomials with positive coefficients and maximum degree $h$. By \eqref{eq:matC}, \eqref{eq:lam}, the inductive hypothesis~\eqref{eq:ind-hyp}, and \Cref{theorem:rankone}, we conclude that for any $a \in \cA_{j^*}$,
    \begin{equation*}
        \vx^{(t)}_i[(j^*, a)] = \sum_{k=1}^{m} \frac{p_{a, k}(\vX_i^{(t)})}{q_{a, k} (\vX_i^{(t)})},
    \end{equation*}
    where $\{p_{a, k}\}, \{q_{a, k}\}$ are multivariate polynomials with positive coefficients and maximum degree $h + 2 |\cA_{j^*}| \leq h + 2 |\cA_i|$. This concludes the inductive step, and the proof.
\end{proof}

Before we proceed, we note that while \Cref{prop:rvu-sigma,prop:rvu-tri} and \Cref{lemma:q-mul-stab,lemma:l-mul-sta} were stated for the construction relating to trigger deviations, those results readily apply for the construction relating to \emph{coarse} trigger deviations as well; we omit the formal statements as they are almost identical to \Cref{prop:rvu-sigma,prop:rvu-tri} and \Cref{lemma:q-mul-stab,lemma:l-mul-sta}.

In this context, combining \Cref{prop:fp-efcce,prop:fp-efce} with \Cref{theorem:rational} we arrive at the following conclusions.

\begin{lemma}
    \label{lemma:gamma-efcce}
    For any parameters $\eta \leq \frac{1}{512 \|\cQ_i\|_1 \depth_i}$, $\etatri \leq \frac{1}{1024 |\Sigma_i|\depth_i}$, and time $t \in \range{T-1}$,
    \begin{align*}
        \|\vx_i^{(t+1)} - \vx_i^{(t)}\|_1 \leq 8 \|\cQ_i\|_1 \depth_i M(\vX_i^{(t)}),
    \end{align*}
    where $M(\vX_i^{(t)})$ is defined as 
    \begin{equation*}
        \max\left\{ \max_{j \in \cJ_i} \left| 1 - \frac{\vlam_i^{(t+1)}[j]}{\vlam_i^{(t)}[j]}\right|, \max_{j  \in \cJ_i} \max_{\sigma \in \Sigma_j} \left| 1 - \frac{\vq^{(t+1)}_{j}[\sigma]}{\vq^{(t)}_{j}[\sigma]} \right| \right\}.
    \end{equation*}
\end{lemma}

\begin{proof}
By \Cref{prop:fp-efcce}, it follows that the fixed point $\Vec{x}_i^{(t)}$ can be expressed, for any $\sigma \in \Sigma_i$, as
\begin{equation*}
    \Vec{x}_i^{(t)}[\sigma] = \sum_{k=1}^m \frac{p_{\sigma, k}\left(\vlam_i^{(t)}, (\vq^{(t)}_{j})_{j \in \cJ_i}  \right)}{q_{\sigma, k}\left(\vlam_i^{(t)}, (\vq^{(t)}_{j})_{j \in \cJ_i} \right)},
\end{equation*}
such that $\{p_{\sigma, k}\}, \{q_{\sigma, k}\}$ are multivariate polynomials in $\vX_i^{(t)} = (\vlam_i^{(t)}, (\vq^{(t)}_{j})_{j \in \cJ_i})$ with positive coefficients and maximum degree $\deg_i \defeq 2 \depth_i$. As a result, similarly to \Cref{lemma:q-mul-stab,lemma:l-mul-sta}, it follows that
\begin{equation*}
    \max_{\sigma \in \Sigma_j} \left| 1 - \frac{\vq_{j}^{(t+1)}[\sigma]}{\vq_{j}^{(t)}[\sigma]} \right| \leq 100 \eta \|\cQ_i\|_1 \leq \frac{100}{256\deg_i},
\end{equation*}
for any $j \in \cJ_i$, and
\begin{equation*}
        \max_{j \in \cJ_i} \left| 1 - \frac{\vlam_i^{(t+1)}[j]}{\vlam_i^{(t)}[j]} \right| \leq 200 \etatri |\Sigma_i| \leq \frac{100}{256\deg_i}.
\end{equation*}
As a result, the claim follows from \Cref{theorem:rational}.
\end{proof}

\gammaefce*

\begin{proof}
By \Cref{prop:fp-efce}, it follows that the fixed point $\Vec{x}_i^{(t)}$ can be expressed, for any $\sigma \in \Sigma_i$, as
\begin{equation*}
    \Vec{x}_i^{(t)}[\sigma] = \sum_{k=1}^m \frac{p_{\sigma, k}\left(\vlam_i^{(t)}, (\vq^{(t)}_{\hat{\sigma}})_{\hat{\sigma} \in \Sigma_i^*}  \right)}{q_{\sigma, k}\left(\vlam_i^{(t)}, (\vq^{(t)}_{\hat{\sigma}})_{\hat{\sigma} \in \Sigma_i^*} \right)},
\end{equation*}
such that $\{p_{\sigma, k}\}, \{q_{\sigma, k}\}$ are multivariate polynomials in $\vX_i^{(t)} = (\vlam_i^{(t)}, (\vq^{(t)}_{\hat{\sigma}})_{\hat{\sigma} \in \Sigma_i^*})$ with positive coefficients and maximum degree $\deg_i \defeq 2 |\cA_i| \depth_i$. As a result, in light of \Cref{lemma:q-mul-stab,lemma:l-mul-sta}, 
\begin{equation*}
    \max_{\sigma \in \Sigma_j} \left| 1 - \frac{\vq_{\hat{\sigma}}^{(t+1)}[\sigma]}{\vq_{\hat{\sigma}}^{(t)}[\sigma]} \right| \leq 100 \eta \|\cQ_i\|_1 \leq \frac{100}{256\deg_i},
\end{equation*}
for any $\hat{\sigma} = (j,a) \in \Sigma_i^*$, and
\begin{equation*}
        \max_{\hat{\sigma} \in \Sigma_i^*} \left| 1 - \frac{\vlam_i^{(t+1)}[\hat{\sigma}]}{\vlam_i^{(t)}[\hat{\sigma}]} \right| \leq 200 \etatri |\Sigma_i| \leq \frac{100}{256\deg_i}.
\end{equation*}
As a result, the claim follows from \Cref{theorem:rational}.
\end{proof}

\subsection{Completing the Proof}

Finally, here we combine all of the previous ingredients to complete the proof of \Cref{cor:trigger}.

\begin{proposition}
    \label{prop:aux}
    Let $\eta \leq \frac{1}{256|\Sigma_i|^{1.5}}$ and $\etatri \leq \frac{1}{512|\Sigma_i|^{2.5}}$. Then, for any $T \geq 2$,
    \begin{align*}
        \max\{0, \reg_{\Psi_i}^T\} \leq \frac{2|\Sigma_i|^2 \log T}{\eta} + \frac{2|\Sigma_i| \log T}{\etatri} + (32\eta |\Sigma_i| |\cQ_i|^2 + 256 \etatri |\Sigma_i|^4)& \sum_{t=1}^{T-1} \|\ut_i^{(t+1)} - \ut_i^{(t)}\|_\infty^2 \\
        +(32\eta |\Sigma_i| |\cQ_i|^2 + 256 \etatri |\Sigma_i|^4) \sum_{t=1}^{T-1} \|\vx_i^{(t+1)} - \vx_i^{(t)}\|_\infty^2 - \frac{1}{512\etatri} \sum_{t=1}^{T-1} &\|\vlam_i^{(t+1)} - \vlam_i^{(t)}\|^2_{\vlam_i^{(t)}, \infty}
        \\-\frac{1}{1024\eta} \sum_{\hat{\sigma} \in \Sigma^*_i} \sum_{t=1}^{T-1} &\| \vq_{\hat{\sigma}}^{(t+1)} - \vq_{\hat{\sigma}}^{(t)}\|^2_{\vq_{\hat\sigma}^{(t)}, \infty}.
    \end{align*}
\end{proposition}

\begin{proof}
    Fix any $t \in \range{T-1}$. By definition of $\Ut_i^{(t)}$ (\Cref{line:outer}),
    \begin{align}
        \|\Ut_i^{(t+1)} - \Ut_i^{(t)}\|^2_\infty &\leq \|\ut_i^{(t+1)} \otimes \vx_i^{(t+1)} - \ut_i^{(t)} \otimes \vx_i^{(t)} \|^2_\infty \notag \\
        &\leq 2 \| \ut_i^{(t+1)} \otimes (\vx_i^{(t+1)} - \vx_i^{(t)}) \|_\infty^2 +2 \| \vx_i^{(t)} \otimes (\ut_i^{(t+1)} - \ut_i^{(t)}) \|_\infty^2 \label{eq:young-tri} \\
        &= 2 \| \ut_i^{(t+1)}\|_\infty^2 \|\vx_i^{(t+1)} - \vx_i^{(t)}\|^2_\infty + 2 \|\vx_i^{(t)}\|_\infty^2 \|\ut_i^{(t+1)} - \ut_i^{(t)}\|^2_\infty \label{eq:inftyflat} \\
        &\leq 2 \|\vx_i^{(t+1)} - \vx_i^{(t)}\|^2_\infty + 2 \|\ut_i^{(t+1)} - \ut_i^{(t)}\|^2_\infty, \label{eq:finbound}
    \end{align}
    where \eqref{eq:young-tri} follows from the triangle inequality for the $\|\cdot\|_\infty$ norm, as well as Young's inequality; \eqref{eq:inftyflat} uses the fact that $\| \vx \otimes \ut \|_\infty = \|\vx\|_\infty \|\ut\|_\infty$, for any vectors $\vx, \ut$; and \eqref{eq:finbound} follows from the assumption that $\|\ut_i^{(t)}\|_\infty, \|\vx_i^{(t)}\|_\infty \leq 1$.
Similarly, for any $t \in \range{T-1}$,
\begin{align}
    \|\uttri^{(t+1)} - \uttri^{(t)}\|^2_\infty &= | \langle \vX_{\hat{\sigma}}^{(t+1)}, \Ut_i^{(t+1)} \rangle - \langle \vX_{\hat{\sigma}}^{(t)}, \Ut_i^{(t)} \rangle |^2 \label{align:deftri} \\
    &\leq 2 | \langle \vX^{(t+1)}_{\hat{\sigma}}, \Ut_i^{(t+1)} - \Ut_i^{(t)} \rangle |^2 + 2 | \langle \Ut_i^{(t)}, \vX_{\hat{\sigma}}^{(t+1)} - \vX_{\hat{\sigma}}^{(t)} \rangle|^2 \label{align:triyoung} \\
    &\leq 8 |\Sigma_i|^2 \|\Ut_i^{(t+1)} - \Ut_i^{(t+1)}\|_\infty^2 +2\|\vq_{\hat{\sigma}}^{(t+1)} - \vq^{(t)}_{\hat{\sigma}}\|_1^2, \label{align:cauchy}
\end{align}
for some $\hat{\sigma} \in \Sigma_i^*$, where \eqref{align:deftri} follows from the definition of $\uttri^{(t)}$; \eqref{align:triyoung} uses Young's inequality; and \eqref{align:cauchy} uses the Cauchy-Schwarz inequality, along with the fact that $\|\Ut_i^{(t)}\|_\infty \leq 1$ and $\|\vX^{(t)}_{\hat{\sigma}}\|_1 \leq 2 |\Sigma_i|$. Further, for any $\hat{\sigma} \in \Sigma_i^*$, $\eta \leq \frac{1}{256|\Sigma_i|^{1.5}}$ and $\etatri \leq \frac{1}{512|\Sigma_i|^{2.5}}$,
\begin{align}
    - \frac{1}{1024\eta} \|\vq_{\hat{\sigma}}^{(t+1)} - \vq_{\hat{\sigma}}^{(t)} \|_{\vq^{(t)}_{\hat{\sigma}}, \infty}^2 + 32 \etatri |\Sigma_i|^2 \|\vq_{\hat{\sigma}}^{(t+1)} - \vq_{\hat{\sigma}}^{(t)} \|_1^2
    &\leq \left( - \frac{1}{1024\eta} + 32 \etatri |\Sigma_i|^4 \right) \|\vq_{\hat{\sigma}}^{(t+1)} - \vq_{\hat{\sigma}}^{(t)} \|_{\infty}^2 \notag \\ 
    &\leq \left( - \frac{|\Sigma_i|^{1.5}}{4} + \frac{|\Sigma_i|^{1.5}}{16} \right) \|\vq_{\hat{\sigma}}^{(t+1)} - \vq_{\hat{\sigma}}^{(t)} \|_{\infty}^2 \notag \\
    &\leq 0. \label{eq:qgone}
\end{align}
As a result, the proof follows from \Cref{prop:rvu-sigma,prop:rvu-tri,prop:rvu-hull}, \eqref{eq:finbound}, \eqref{align:cauchy}, and \eqref{eq:qgone}.
\end{proof}

As a result, we are now ready to establish \Cref{cor:rvu}, the statement of which is recalled below.

\corrvu*

\begin{proof}
    By \Cref{lemma:gamma-efce},
    \begin{align*}
        \frac{1}{512\etatri} \sum_{t=1}^{T-1} \|\vlam_i^{(t+1)} - \vlam_i^{(t)}\|^2_{\vlam_i^{(t)}, \infty} + \frac{1}{1024\eta} \sum_{\hat{\sigma} \in \Sigma^*_i} \sum_{t=1}^{T-1} &\| \vq_{\hat{\sigma}}^{(t+1)} - \vq_{\hat{\sigma}}^{(t)}\|^2_{\vq_{\hat\sigma}^{(t)}, \infty} \\
        \geq &\frac{1}{2^{14} \eta \|\cQ_i\|^2_1 \deg^2_i} \|\vx_i^{(t+1)} - \vx_i^{(t)}\|_1^2.
    \end{align*}
    Thus, the proof follows directly from \Cref{prop:aux} since for any $t \in \range{T-1}$,
    \begin{equation*}
        \left( 256 \eta |\Sigma_i|^3 - \frac{1}{2^{15}\eta \|\cQ_i\|_1^2 \deg_i^2} \right) \|\vx_i^{(t+1)} - \vx_i^{(t)}\|_1^2 \leq 0,
    \end{equation*}
    for any $\eta \leq \frac{1}{2^{12}|\Sigma_i|^{1.5}\|\cQ_i\|_1 \deg_i}$.
\end{proof}

So far we have performed the analysis from the perspective of a fixed player $i \in \range{n}$, while being oblivious to the mechanism that produces the sequence of utilities $(\ut_i^{(t)})_{1 \leq t \leq T}$. Having established the RVU bound of \Cref{cor:rvu}, we are ready to show that when \emph{all} players employ our learning dynamics, the second-order path length is bounded by $O(\log T)$. (In what follows, we tacitly assume that each player uses $\etatri \defeq \frac{1}{2|\Sigma_i|} \eta$, in accordance with \Cref{cor:rvu}.)

\begin{theorem}
    \label{cor:pathlength}
    Suppose that each player $i \in \range{n}$ uses learning rate $\eta \leq \frac{1}{2^{12} (n-1) |\Sigma|^{1.5} \|\cQ\|_1 |\cZ| \deg }$, where $\deg = 2 |\cA| \depth$. Then, for any $T \geq 2$,
    \begin{equation*}
        \sum_{t=1}^{T-1} \sum_{i=1}^n \|\vx_i^{(t+1)} - \vx_i^{(t)}\|_1^2 \leq 2^{19} n |\Sigma|^2 \|\cQ\|_1^2 \deg^2 \log T.
    \end{equation*}
\end{theorem}

\begin{proof}
    For any time $t \in \range{T-1}$ and player $i \in \range{n}$,
    \begin{equation*}
        \|\ut_i^{(t+1)} - \ut_i^{(t)}\|_\infty^2 \leq (n-1) |\cZ|^2 \sum_{i' \neq i} \|\vx_{i'}^{(t+1)} - \vx_{i'}^{(t)}\|_1^2,
    \end{equation*}
    by~\citep[Claim 4.16]{Anagnostides22:Faster}. Further, for $\eta \leq \frac{1}{2^{12} (n-1) |\Sigma|^{1.5} \|\cQ\|_1 |\cZ| \deg }$, 
    \begin{equation*}
        \left( 256 \eta (n-1)^2 |\Sigma|^3 |\cZ|^2 - \frac{1}{2^{16}\eta \|\cQ\|^2_1 \deg^2} \right) \leq 0.
    \end{equation*}
    As a result, using \Cref{cor:rvu}, $\sum_{i=1}^n \max\{0, \reg^T_{\Psi_i}\}$ can be upper bounded by
    \begin{equation*}
        \frac{8 n |\Sigma|^2 \log T}{\eta} - \frac{1}{2^{16}\eta \deg^2 \|\cQ\|_1^2} \sum_{i=1}^n \sum_{t=1}^{T-1} \|\vx_i^{(t+1)} - \vx_i^{(t)}\|_1^2.
    \end{equation*}
    But, given that $\sum_{i=1}^n \max\{0, \reg^T_{\Psi_i}\} \geq 0$, we conclude that
    \begin{equation*}
        \sum_{i=1}^n \sum_{t=1}^{T-1} \|\vx_i^{(t+1)} - \vx_i^{(t)}\|_1^2 \leq 2^{19} n |\Sigma|^2 \deg^2 \|\cQ\|_1^2 \log T.
    \end{equation*}
\end{proof}

We now arrive at \Cref{cor:trigger}, which is restated below with the precise parameterization. 

\begin{corollary}
    \label{cor:precise-efce}
Suppose that all players employ \Cref{algo:Gordon} instantiated with $\algoshort$ for all local regret minimizers, $\regdep_{\triangle}$ and $\{\regdep_{\hat{\sigma}}\}_{\hat{\sigma} \in \Sigma_i^*}$, with $\eta = \frac{1}{2^{13} (n-1) |\Sigma|^{1.5} \|\cQ\|_1 |\cZ| |\cA| \depth } $ and $\etatri = \frac{1}{2|\Sigma_i|} \eta$. Then, the trigger regret of each player $i \in \range{n}$ after $T$ repetitions will be bounded as
        \begin{equation*}
        \reg^T_{\Psi_i} \leq 2^{17} n |\Sigma|^{3.5} \|\cQ\|_1 |\cZ| |\cA| \depth \log T.
    \end{equation*}
\end{corollary}

\begin{proof}
    This follows directly from \Cref{cor:rvu} and \Cref{cor:pathlength}. 
\end{proof}

\begin{corollary}
    \label{cor:efcce}
    Suppose that all players employ \Cref{algo:Gordon} instantiated with $\algoshort$ for all local regret minimizers, $\regdep_{\triangle}$ and $\{\regdep_{j}\}_{j \in \cJ_i}$, with $\eta = \frac{1}{2^{13} (n-1) |\Sigma|^{1.5} \|\cQ\|_1 |\cZ| \depth } $ and $\etatri = \frac{1}{2|\Sigma_i|} \eta$. Then, the trigger regret of each player $i \in \range{n}$ after $T$ repetitions will be bounded as
        \begin{equation}
            \label{eq:loose}
        \reg^T_{\Psi_i} \leq 2^{17} n |\Sigma|^{3.5} \|\cQ\|_1 |\cZ| \depth \log T.
    \end{equation}
\end{corollary}

\begin{proof}
    The proof is analogous to \Cref{cor:precise-efce}.
\end{proof}

We remark that for \emph{coarse} trigger regret, our bound~\eqref{eq:loose} is loose, as the analysis is not optimized to handle coarse trigger deviation functions; instead, \Cref{cor:efcce} follows the construction of trigger deviations, with the exception of using \Cref{lemma:gamma-efcce} in order to obtain a slightly improved RVU bound. Further refining \Cref{cor:efcce} was not within our scope.
\section{Description of the Game Instances}
\label{appendix:games}

In this section, to keep our paper self-contained, we describe the games we used in our experiments (\Cref{section:experiments}), as well as the precise parameterization for each instance.

\paragraph{Kuhn poker} \emph{Kuhn poker} is a simple poker variant studied by~\citet{Kuhn53:Extensive}. For simplicity, below we describe the $2$-player version of Kuhn poker; the $3$-player version we consider in our experiments is analogous. 

In Kuhn poker each player initially submits an ante worth of $1$ in the pot. Then, each player is privately dealt one card from a deck of $r$ unique cards---or \emph{ranks}; in our experiments we used $r = 3$. Next, a single round of betting occurs: First, player $1$ gets to decide either check or bet. Then, 
\begin{itemize}
    \item If player $1$ checked, the second player can either check or raise.
    \begin{itemize}
        \item If player $2$ also checked, a ``showdown'' occurs, meaning that the player with the highest card wins the pot, thereby terminating the game. 
        \item On the other hand, if player $2$ raised, player $1$ can either fold or call; in the former case player $2$ wins the pot, while in the latter a showdown follows.
    \end{itemize}
    \item If player $1$ raised, player $2$ can either fold or call.
    \begin{itemize}
        \item If player $2$ folded, then player $1$ wins the pot, while
        \item if player $2$ called, a showdown occurs.
    \end{itemize}
\end{itemize}

\paragraph{Sheriff} \emph{Sheriff}~\citep{Farina19:Correlation} is a $2$-player bargaining game inspired by the board game ``Sheriff of Nottingham.'' Initially, player $1$ (or the ``Smuggler'') secretly loads his cargo with $m \in \{0, 1, \dots, m_{\text{max}}\}$ illegal items. The game then proceeds for $r$ bargaining rounds. In each round,
\begin{itemize}
    \item the Smuggler first gets to decide a bribe amount $b$ in $\{0, 1, \dots, b_{\text{max}}\}$. This amount also becomes available to player $2$ (the ``Sheriff''), although the smuggler does not transfer than amount unless it is the ultimate round.
    \item The Sheriff then decides whether to accept the bribe. 
    \begin{itemize}
        \item If the Sheriff accepts the bribe of value $b$, the smuggler gets a payoff of $p \cdot m - b$, while Sheriff receives a payoff of $b$.  
        \item In the contrary case, Sheriff decides whether to inspect the cargo. 
        \begin{itemize}
            \item If the Sheriff does not inspect the cargo, the Smuggler receives a payoff of $v \cdot m$, while the Sheriff gets $0$ utility; 
            \item Otherwise, if the Sheriff detects illegal items, the Smuggler must pay the Sheriff an amount of $p \cdot m$, while if no illegal items were loaded, the Sheriff has to compensate the Smuggler with a utility of $s$.
        \end{itemize}
    \end{itemize}
\end{itemize}

In our experiments, we use the baseline version of Sheriff, wherein $v = 5, p = 1, s = 1, m_{\text{max}} = 5, b_{\text{max}} = 2$, and $r = 2$.

\paragraph{Goofspiel} Goofspiel is a $2$-player card game introduced by~\citet{Ross71:Goofspiel}. The game is based on three identical decks of $r$ cards each, with values ranging from $1$ to $r$; we use $r = 3$ in our experiments. Initially, each player is dealt a full deck, while the third deck (the ``prize'' deck) is faced down on the board after being shuffled. In each round, the topmost card from the prize deck is revealed. Then, each player privately selects a card from their hand with the goal of winning the card that was revealed from the prize deck. The players' selected cards are revealed simultaneously, and the card with the highest value prevails; in case of a tie, the prize card is discarded. This tie-breaking mechanism makes the game general-sum. Finally, the score of each player is the sum of the values of the prize cards that player has won.
\fi
\end{document}